\documentclass[useAMS,usenatbib,psfig,psfig]{mn2e}
\usepackage{graphicx}

\newcommand{\HII}{H {\small II}  }
\newcommand{\kms}{{\rm ~km~s}^{-1}}
\newcommand{\Msun} {M_\odot}

\newcommand{\cmcub}{cm$^{-3}$}

\title[Variability in UC and HC HII regions]{Time variability in simulated ultracompact and hypercompact HII regions}

\author[R.~Galv\'an-Madrid et al.]
{\parbox{\textwidth}{Roberto Galv\'an-Madrid$^{1,2,3}$\thanks{E-mail: \texttt{rgalvan@cfa.harvard.edu}}, 
Thomas Peters$^{1,4}$\thanks{E-mail: \texttt{thomas.peters@ita.uni-heidelberg.de}}, 
Eric R. Keto$^{1}$, 
Mordecai-Mark Mac Low$^{5}$, 
Robi Banerjee$^{4}$,  and 
Ralf S. Klessen$^{4}$
}\vspace{0.4cm}\\
\parbox{\textwidth}{$^{1}$Harvard-Smithsonian Center for Astrophysics, 60 Garden Street, Cambridge MA 02138, USA\\
$^{2}$Centro de Radioastronom\'ia y Astrof\'isica, Universidad Nacional Aut\'onoma de M\'exico, Morelia 58090, Mexico\\
$^{3}$Academia Sinica Institute of Astronomy and Astrophysics, Taipei 106, Taiwan\\
$^{4}$Zentrum f\"{u}r Astronomie der Universit\"{a}t Heidelberg, 
Institut f\"{u}r Theoretische Astrophysik, 
Albert-Ueberle-Str. 2, D-69120 Heidelberg, Germany\\
$^{5}$Department of Astrophysics, American Museum of Natural History,
79th Street at Central Park West, New York, New York 10024-5192, USA\\
}}

\begin{document}

\date{MNRAS in press}

\pagerange{\pageref{firstpage}--\pageref{lastpage}} \pubyear{2011}

\maketitle

\label{firstpage}

\begin{abstract}

Ultracompact and hypercompact \HII regions appear when a star
with a mass larger than about 15 solar masses starts to ionize its own
environment. Recent observations of time variability in these objects are
one of the pieces of evidence that suggest that at least some of them
harbor stars that are still accreting from an infalling
neutral accretion flow that becomes ionized in its innermost part.
We present an analysis of the properties of the \HII regions formed
in the 3D radiation-hydrodynamic simulations presented by Peters et al. as a function of time. 
Flickering of the \HII regions is a natural outcome of this model. 
The radio-continuum fluxes  of the simulated \HII regions, as well as their flux and size variations
are in agreement with the available observations. 
From the simulations, we estimate that 
a small but non-negligible fraction ($\sim 10~\%$) of observed \HII regions should have 
detectable flux variations (larger than $10~\%$) on timescales of $\sim 10$ years, 
with positive variations being more likely to happen than negative variations. 
A novel result of these simulations is that negative flux changes do happen, in contrast 
to the simple expectation of ever growing \HII regions.
We also explore the temporal correlations
between properties that are directly observed (flux and size) and other quantities 
like density and ionization rates.

\end{abstract}


\section{Introduction}

\indent

The most massive stars in the Galaxy, O-type stars with masses $M_\star>20~\Msun$, 
emit copious amounts of UV photons \citep{Vacca96} that ionize part of the dense gas 
from which they form. The resulting \HII regions 
are visible via their free-free continuum and recombination line radiation 
\citep{MezHen67,WC89}. \HII regions span orders of magnitude in size, from 
giant ($D \sim 100$ pc) bubbles, to ``ultracompact" (UC)
and ``hypercompact" (HC) \HII regions, loosely defined as those with sizes of  
$\sim 0.1$ pc and $\sim 0.01$ pc (or less), respectively 
\citep[see the reviews by][]{Church02,Kurtz05,Hoare07}.  
UC and HC \HII regions are the most deeply embedded, and so are best observed at radio 
wavelengths. 

Large, rarefied \HII regions expand  
without interruption within the surrounding medium due to the high pressure contrast 
between the ionized and neutral phases \citep{Spitzer78}. 
This simple model was extrapolated to the ever smaller objects recognized later   
and is widely used to interpret observations of UC and HC \HII regions. 
Common assumptions about these objects are: i) They are steadily expanding within their 
surrounding medium at the sound speed of the ionized gas, $\sim 10$ $\kms$. 
ii) The ionizing star(s) is already formed, i.e., accretion to the massive star(s) 
powering the \HII region has stopped.  

However, evidence has accumulated that suggests a revision of these assumptions: 
\begin{enumerate}
\item 
The ``hot molecular cores" embedding UC and HC \HII regions are 
often rotating and infalling 
\citep[e.g.,][]{Keto90,Cesa98,Beltran06,Beltran10,Klaass09}, 
sometimes from parsec scales all the way to the immediate surroundings of the ionized region 
\citep{GM09,Baobab10}. 

\item
Infall of gas at velocities of a few $\kms$ directly toward the 
ionized center has also been observed in UC and HC \HII regions \citep{ZH97,Beltran06,GM09}. 

\item
The inner ionized gas has been resolved  in a few cases, and it also shows accretion 
dynamics \citep[outflow, infall, and rotation, e.g.,][]{KW06,Sewi08,GM09}.

\item
The spectral index $\alpha$ (where the flux goes as $S_\nu \propto \nu^\alpha$) of some 
UC and HC \HII regions is $\sim 1$ from cm to mm wavelengths, indicating density 
gradients and/or clumpiness inside the ionized gas \citep{Franco00,Ignace04,Keto08,Avalos09}. 
\item
A few UC and HC \HII regions have been shown to have variations on timescales of 
years \citep{Acord98,FHR04, vdT05,GM08}.  
These variations indicate that UC and HC \HII regions sometimes expand \citep{Acord98}, 
and sometimes contract \citep{GM08}. 
Some other ionized regions around massive protostars 
have been shown to remain approximately constant in flux \citep{Goddi11}.

\end{enumerate}

All these observations strongly suggest that UC and HC \HII regions {\it are not} homogeneous spheres 
of gas freely expanding into a quiescent medium, but rather that these small \HII regions are 
intimately related to the accretion processes forming the massive stars. 
Simple analytic models show that the observed \HII region can be either the ionized, 
inner part of the inflowing accretion flow \citep{Keto02,Keto03}  
or the ionized photoevaporative outflow \citep{Hollen94} fed by accretion \citep{Keto07}.

Numerical simulations of the formation and expansion of \HII regions
in accretion flows around massive stars have only recently become
possible with three-dimensional radiation-hydrodynamics.
Studies that simulate the expansion of \HII regions have
focussed on larger-scale effects on the parental molecular cloud 
\citep{Dale05,Dale07a,Dale07b,Peters08,Grit09,Arthur11}. 
However, in none of those simulations 
was the ionizing radiation produced by the massive
stars dynamically forming through gravitational collapse
in the molecular cloud. Recent simulations by 
\cite{Peters10} (hereafter Paper I) 
include a more realistic treatment of the formation of the star cluster. 
Radio-continuum images
generated from the output of the simulations of Paper I show time
variations in the morphology and flux from the \HII regions produced
by the massive stars in formation. These changes are the result
of the complex interaction of the massive filaments of neutral gas
infalling to the central stars with the ionized regions
produced by some of them. A statistical analysis of the \HII 
region morphologies \citep[][Paper II]{Peters10b} consistently
reproduces the relatively high fraction of spherical and unresolved
regions found in observational surveys. Thus, the non-monotonic expansion of \HII 
regions, or flickering, appears able to resolve the excess number of observed UC and HC \HII regions 
with respect to the expectation if they expand uninterrupted 
\citep[the so-called lifetime problem,][]{WC89}. Furthermore, the ionizing radiation
is unable to stop protostellar growth when accretion is strong enough.
Instead, accretion is stopped by the fragmentation of the
gravitationally unstable accretion flow in a process we call
``fragmentation-induced starvation", a theoretical discussion of which
can be found in \cite{Peters10c} (Paper III).

In this paper we present a more detailed analysis of the flux 
variability in the simulations presented in Paper I.  
In \S 2 we describe the set of numerical simulations and the methods of analysis. 
In \S 3 we present our results. \S 4 discusses the implications of our findings. 
In \S 5 we present our conclusions.

\section{Methods}

\subsection{The Numerical Simulations}

\indent

Our study uses the highest resolution simulations of those presented in Paper I. The simulations 
use a modified version of the adaptive-mesh code FLASH \citep{Fryx00}, including self-gravity 
and radiation feedback. They include for the first time a self-consistent treatment 
of gas heating by both ionizing and non-ionizing radiation. We refer the reader to Paper I for further 
details of the numerical methods. The initial conditions are a cloud mass of 1000 $\Msun$ with an 
initial temperature of 30 K. The initial density distribution is a flat inner region of 0.5 pc 
radius surrounded by a region with a decreasing density $\propto r^{-1.5}$. The density in the 
homogeneous volume is $1.27 \times 10^{-20}$ g \cmcub. The simulation box has a length of 3.89 pc. 
The size and mass of the cloud are in agreement with those of star-formation regions that are 
able to produce at least one star with $M_\star>20~\Msun$  \citep[e.g.,][]{GM09}.

Run A (as labeled in Paper I) has a maximum cell resolution of 98 AU and only the first collapsed 
object is followed as a sink particle \citep[see ][]{Federr10}. In this run, the formation of additional stars 
(the sink particles) is suppressed using a density-dependent temperature floor (see Paper I for details). 
On the other hand, in run B additional collapse events are permitted and a star cluster 
is formed, each star being represented by a sink particle. 
The maximum resolution in Run B is also 98 AU.

\subsection{Data Sets}

\indent

For the entire time span of Run A (single sink) and Run B (multiple sink), 
radio-continuum maps at a wavelength of 2 cm were generated from the simulation output 
every $\sim 300$ yr by integrating the radiative transfer equation for free-free radiation 
while neglecting scattering \citep{GS02}. 
Following \cite{MacLow91}, each intensity map was then convolved to a circular Gaussian beam 
with half-power beam width HPBW $=0.14''$ (assuming a source distance of 2.65 kpc). A noise 
level of $10^{-3}$ Jy was added to each image. Further details are given in Paper I.    
These maps were used to explore the behavior of the free-free continuum from the \HII region 
over the entire time evolution of the simulations. 
For Run B, sometimes the \HII regions overlap both physically in space and/or in appearance 
in the line of sight. 
Therefore, the presented time analysis refers to the entire star cluster unless otherwise 
specified.

To compare more directly to available observations, which span 
at most a couple of decades in time, each of Run A and B were re-run in four time intervals  
for which a flux change was observed in the radio-continuum images mentioned above.  
Data dumps and radio-continuum maps were generated at every simulation time step ($\sim 10$ yr). 
The analysis performed in the low time-resolution maps was also done in these high time-resolution 
data.  
For Run B, the intervals for the high time-resolution data were also selected such that the 
\HII region powered by the most massive star is reasonably isolated 
from fainter \HII regions ionized by neighboring sink particles, 
both in real space and in the synthetic maps. 

\begin{figure}
\includegraphics[width=84mm]{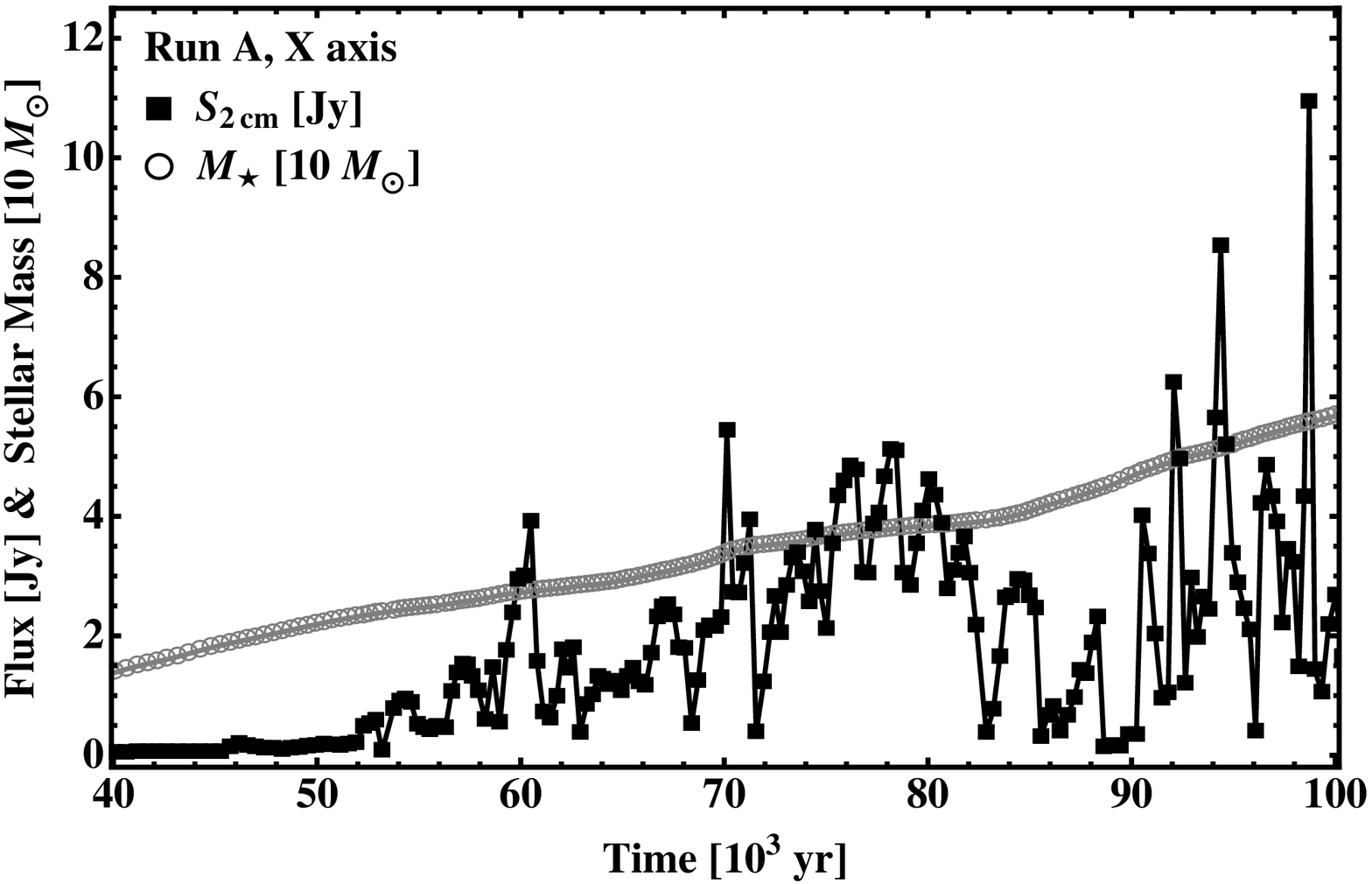}
\includegraphics[width=84mm]{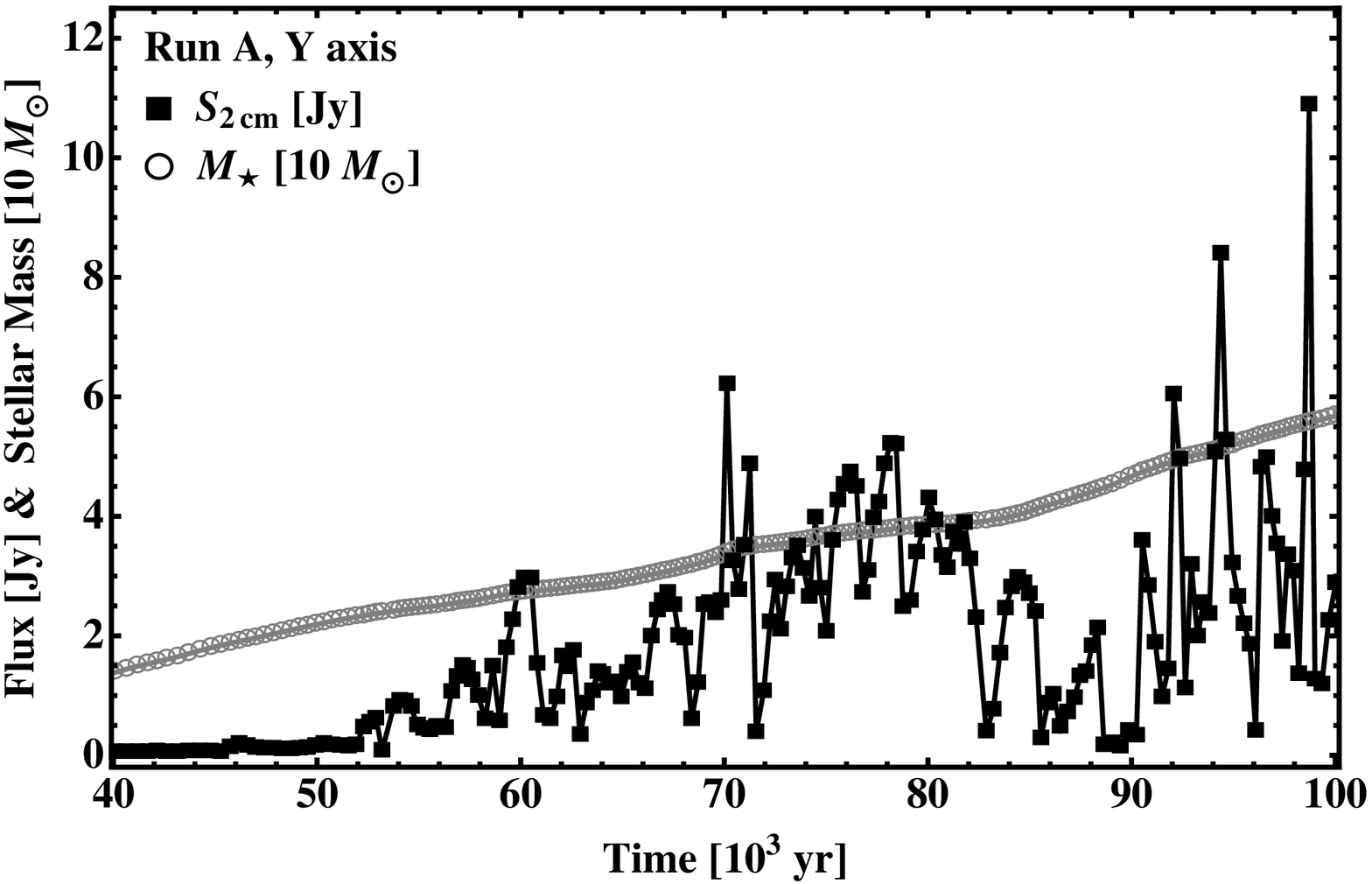}
\includegraphics[width=84mm]{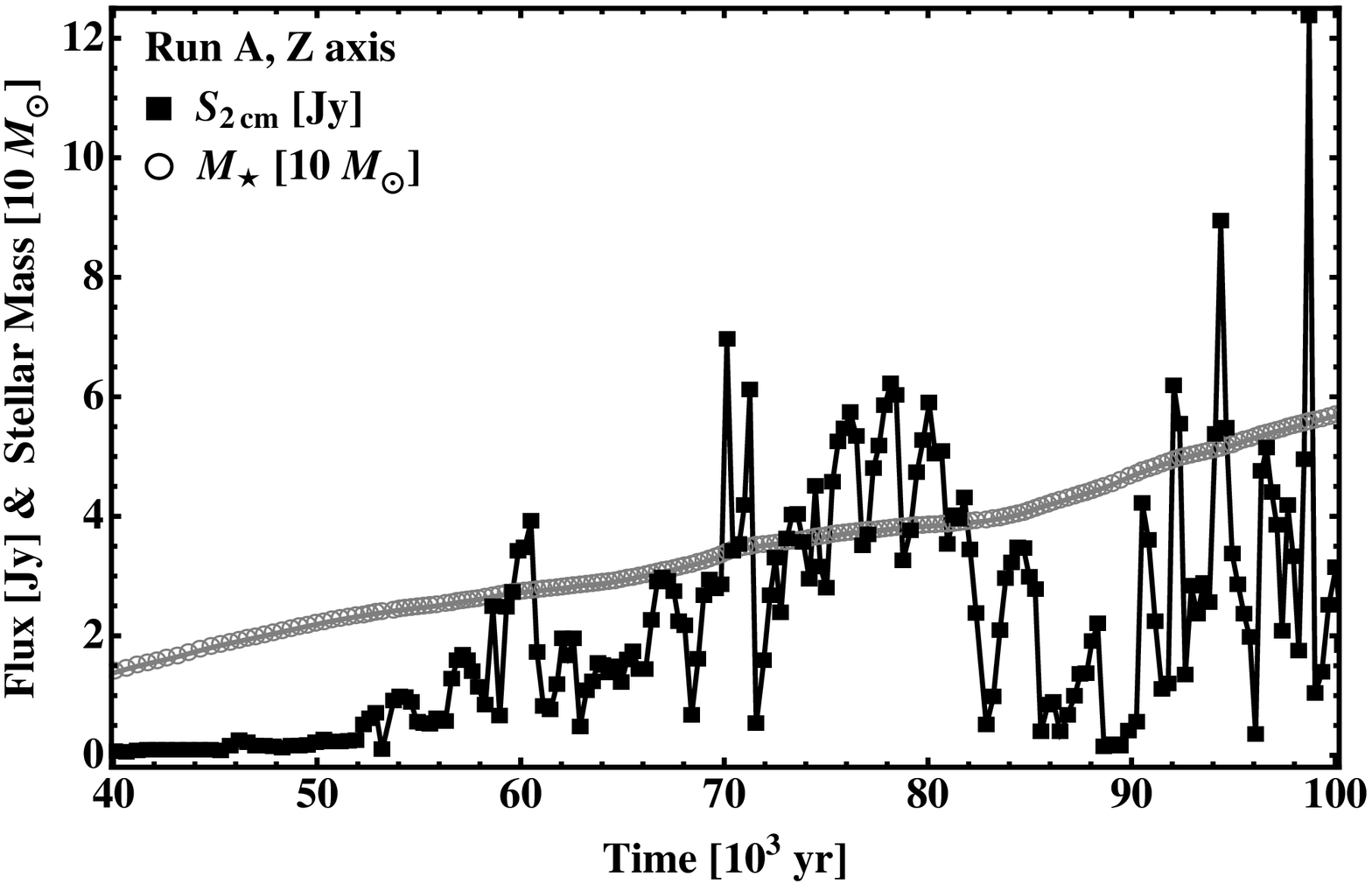}
 \caption{2-cm flux ($S_{\rm 2cm}$, filled black squares) and stellar mass ($M_\star$, 
 gray circles) as a function of time ($t$) for Run A. 
 Although the long-term trend of the \HII region is to increase in flux, it constantly 
 flickers during its evolution. The fluxes at the different projections (X-axis {\it top}, Y-axis 
{\it middle}, Z-axis {\it bottom}), though not the same, follow each other closely.}
  \label{fig1}
\end{figure}

\section{Results}

\subsection{Variable HII Regions}

\indent

Real UC and HC \HII regions, as well as those that result from the simulations presented 
here, are far from the ideal cases \citep[e.g.,][]{Spitzer78}. However, it is instructive 
to discuss the limiting ideal cases to show that their variability is a natural consequence 
of their large to moderate optical depth. 

The flux density $S_\nu$ of an ionization-bounded\footnote{For \HII regions that are embedded in 
their parental cloud, the ionization-bounded approximation is better than the density-bounded 
approximation. The analysis here presented can be derived from, e.g., 
\cite{MezHen67,Spitzer78,RL79}; and \cite{Keto03}.}
\HII region is

\begin{equation}
S
= 
\frac{2 k \nu^2}{c^2} 
\int_\Omega T_\mathrm{B} d\Omega, 
\end{equation}

\noindent
where $k$ is the Boltzmann constant, $c$ is the speed of light, $\nu$ is the frequency, and the 
brightness temperature $T_\mathrm{B}$ is integrated over the angular area $\Omega$ of the \HII region. 

In the limit of very low free-free optical depths 
($\tau_\mathrm{ff} \ll 1$), $T_\mathrm{B}$ along a line of sight $l$ goes as

\begin{equation}
T_\mathrm{B} \propto T_e^{-0.35} \nu^{-2.1} \int_l n^2 dl, 
\end{equation}

\noindent
where $T_e$ and $n$ are the electron temperature and density respectively. 

Combining equations (1) and (2) we have that for a given $\nu$ and constant $T_e$: 

\begin{equation}
S (\tau_\mathrm{ff} \ll 1) \propto n^2 R^3 \propto \dot{N}, 
\end{equation}

\noindent
where the last proportionality comes from the Str\"omgren relation $\dot{N} \propto n^2 R^3$  
($\dot{N}$ is the ionizing-photon rate and $R$ is the `radius' of the \HII region). 
Equation (3) shows that in the optically-thin limit the \HII-region flux only depends 
on $\dot{N}$. For time intervals of a few $\times 10$ yr the mass and
ionizing flux of an accreting protostar remain almost constant, and so does the flux of an  
associated optically-thin \HII region.  

On the other hand, for very high optical depths ($\tau_\mathrm{ff} \gg 1$) $T_\mathrm{B}=T_e$. 
For a given frequency and constant $T_e$, equation (1) becomes: 

\begin{equation}
S (\tau_\mathrm{ff} \gg 1) \propto R^2 \propto \dot{N}^{2/3} n^{-4/3}, 
\end{equation}

\noindent 
therefore, the flux of an optically thick \HII region is proportional to 
its area, and both flux and area decrease with density. 

This analysis is valid for time intervals larger than the recombination timescale 
\citep[$\sim 1$ month for $n \sim 10^6$ cm$^{-3}$, see e.g.,][]{Oster89} and 
as long as the growth of $\dot{N}$ is negligible. 

\smallskip

However, the \HII regions in the simulations are clumpy and have subregions of high and low 
free-free optical depth. Their flux during the accretion stage (while they flicker) 
is dominated by the denser, optically thicker ($\tau_\mathrm{ff} > 1$) subregions, 
so their behaviour is closer to eq. (4) than to eq. (3). 
The on-line version of this paper contains a movie of $\tau_{ff}$ for Run B 
as viewed from the Z-axis (line of sight perpendicular to the plane of the accretion flow).  
The clumpiness and intermediate-to-large optical depth of these \HII regions are also the reasons 
behind their rising 
spectral indices up to relatively large frequencies ($\nu>100$ GHz) without a significant 
contribution from dust emission (see the analytical discussions of Ignace \& Churchwell 2004 and 
Keto et al. 2008, for an analysis of these simulations see Paper II). 
As for the variability, the large optical 
depths cause the size and flux of the simulated \HII regions to be well correlated with each other, 
and anticorrelated with the density of the central ionized gas (see Section 3.6). 
The neutral accretion flow in which the ionizing sources are embedded is filamentary and prone 
to gravitational instability (further discussion is in Section 3 of Paper III, see also Paper II). 
The changes in the density of the \HII regions are a consequence of their passage through 
density enhancements in the quickly evolving accretion flow.

\subsection{Global Temporal Evolution}

\begin{figure}
\includegraphics[width=84mm]{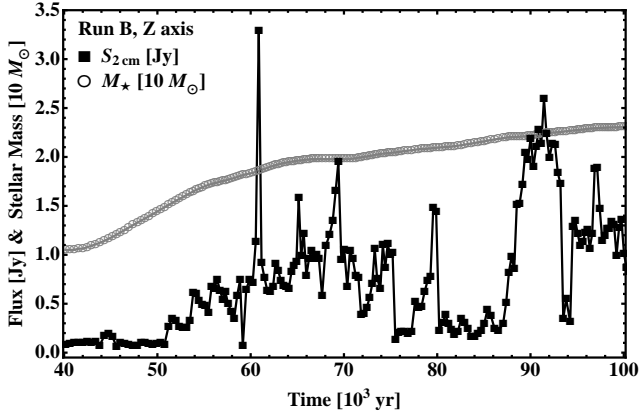}
 \caption{2-cm flux ($S_{\rm 2cm}$, filled black squares) of the \HII region formed by 
 the most massive star and its 
 stellar mass ($M_\star$, gray circles) as a function of time ($t$) for Run B. 
 Although the long-term trend of the \HII region is to increase in flux, it constantly 
 flickers during its evolution. Only the Z-axis projection is used because from this  
viewing angle, perpendicular to the plane of the accretion flow, the brightest \HII region can 
be distinguished from other \HII regions at all times.}
  \label{fig2}
\end{figure}

\begin{figure}
\includegraphics[width=84mm]{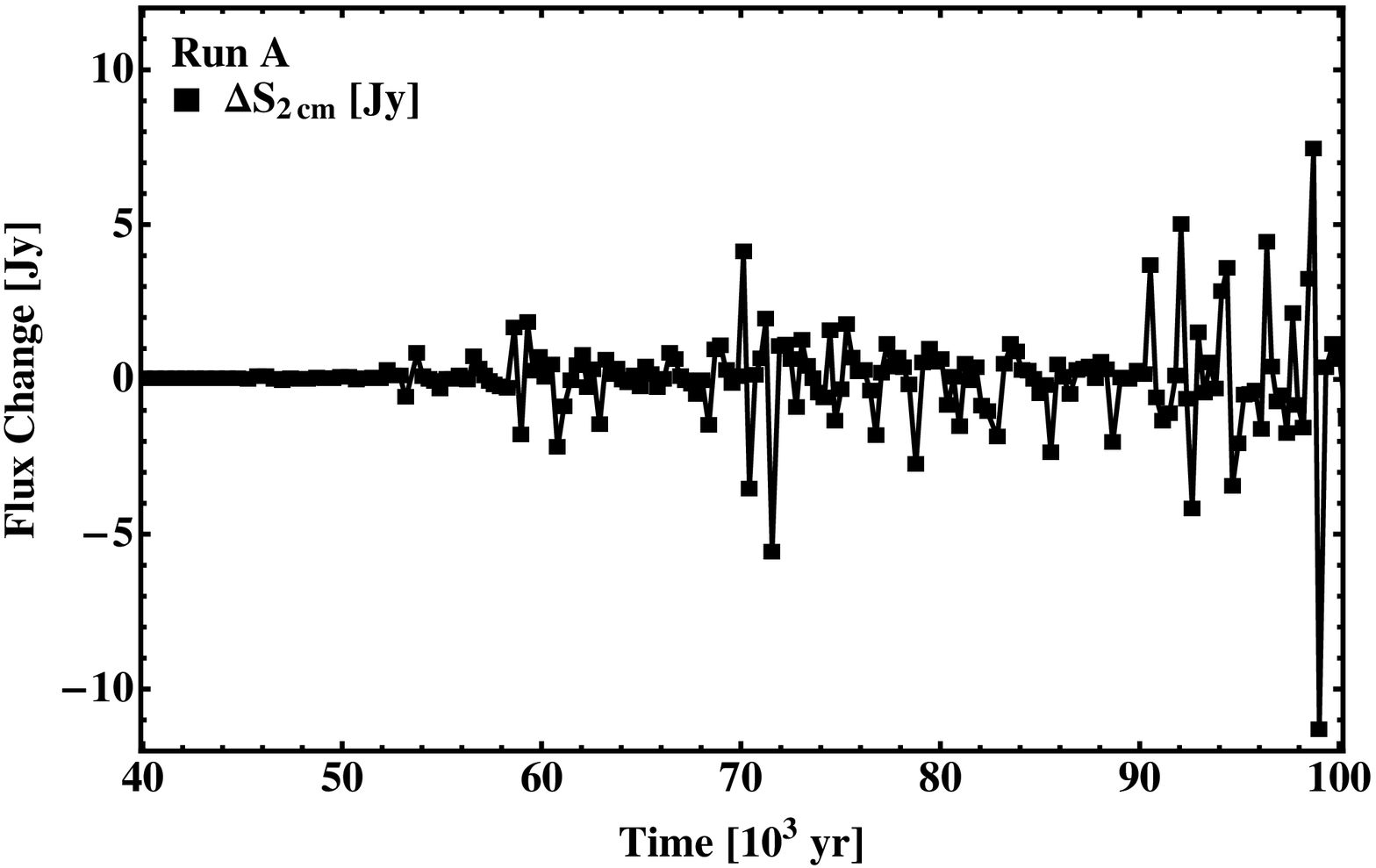}
\includegraphics[width=84mm]{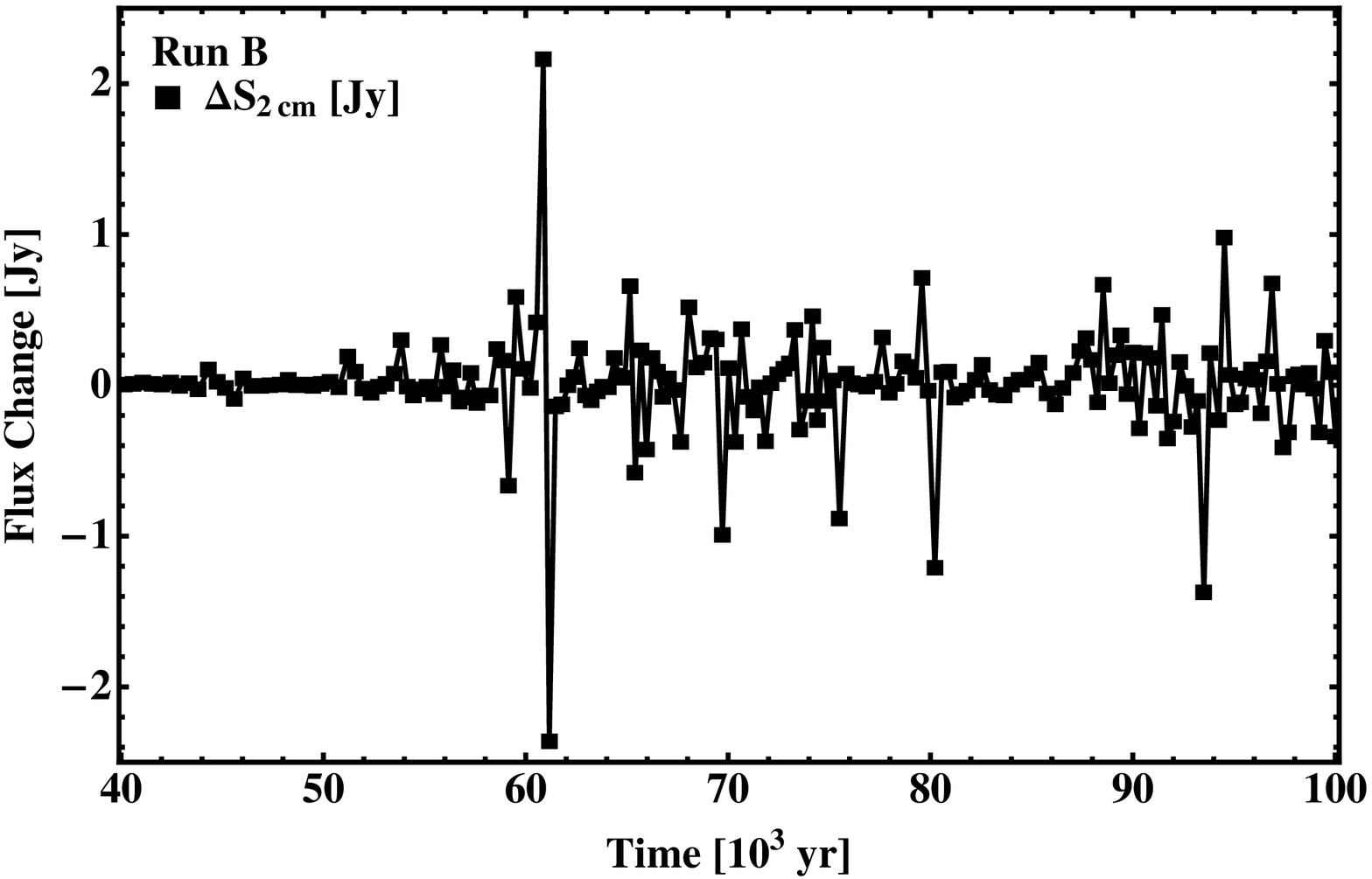}
\caption{2-cm flux variations  ($\Delta S_{\rm 2cm}$) 
for Run A ({\it top} panel) and Run B 
({\it bottom}) panel.} 
\label{fig3}
\end{figure}

\indent

Figure 1 shows the global temporal evolution of the \HII region in Run A (single sink particle). 
The 2-cm flux ($S_{\rm 2cm}$) observed from orthogonal directions 
and  the mass of the ionizing star ($M_\star$) are plotted against time. 
The global temporal trend of the \HII region is to expand and become brighter. 
However, fast temporal variations are seen at all the stages of the evolution. 
The fluxes in the projections along the three different cartesian axes follow each other closely. 
For the rest of the analysis, the Z-axis projection, a line of sight perpendicular to the 
plane of the accretion flow, is used.

\begin{figure}
\includegraphics[width=84mm]{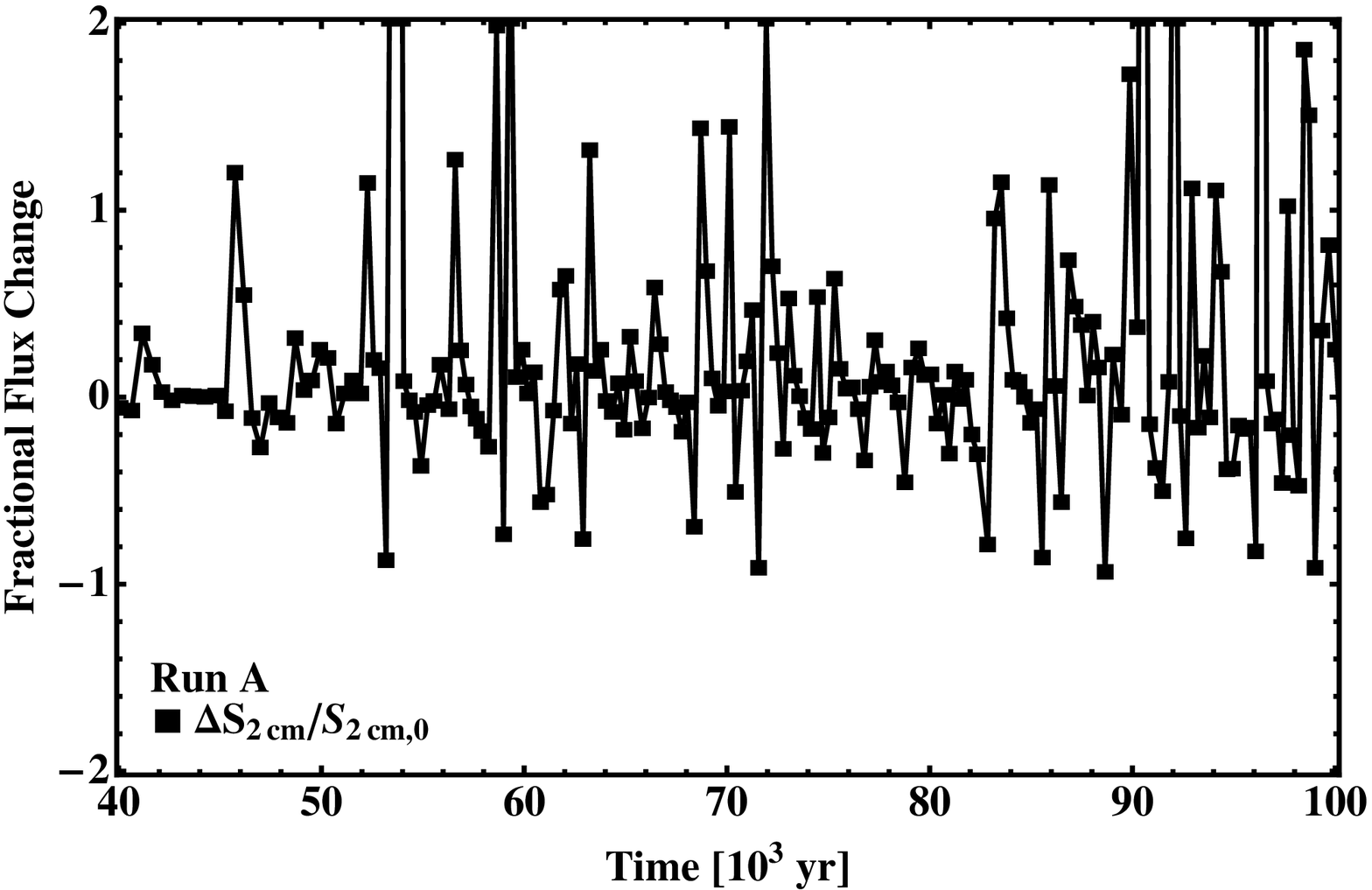}
\includegraphics[width=84mm]{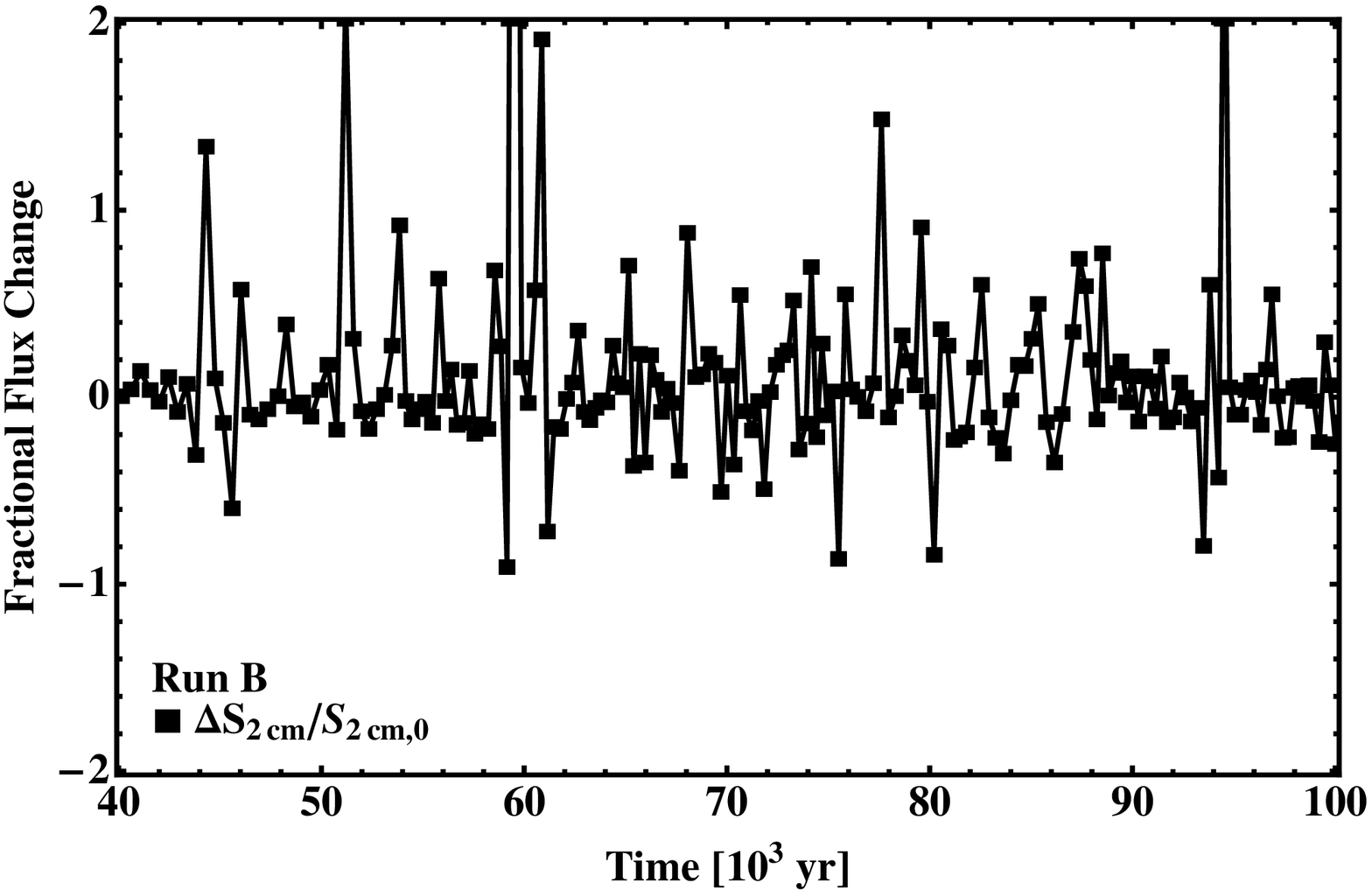}
 \caption{Fractional flux variations
 ($\Delta S_{\rm 2cm} / S_{\rm 2cm,0}$) 
 for Run A ({\it top} panel) and Run B ({\it bottom}) panel.}
  \label{fig4}
\end{figure}

\begin{figure}
\includegraphics[width=84mm]{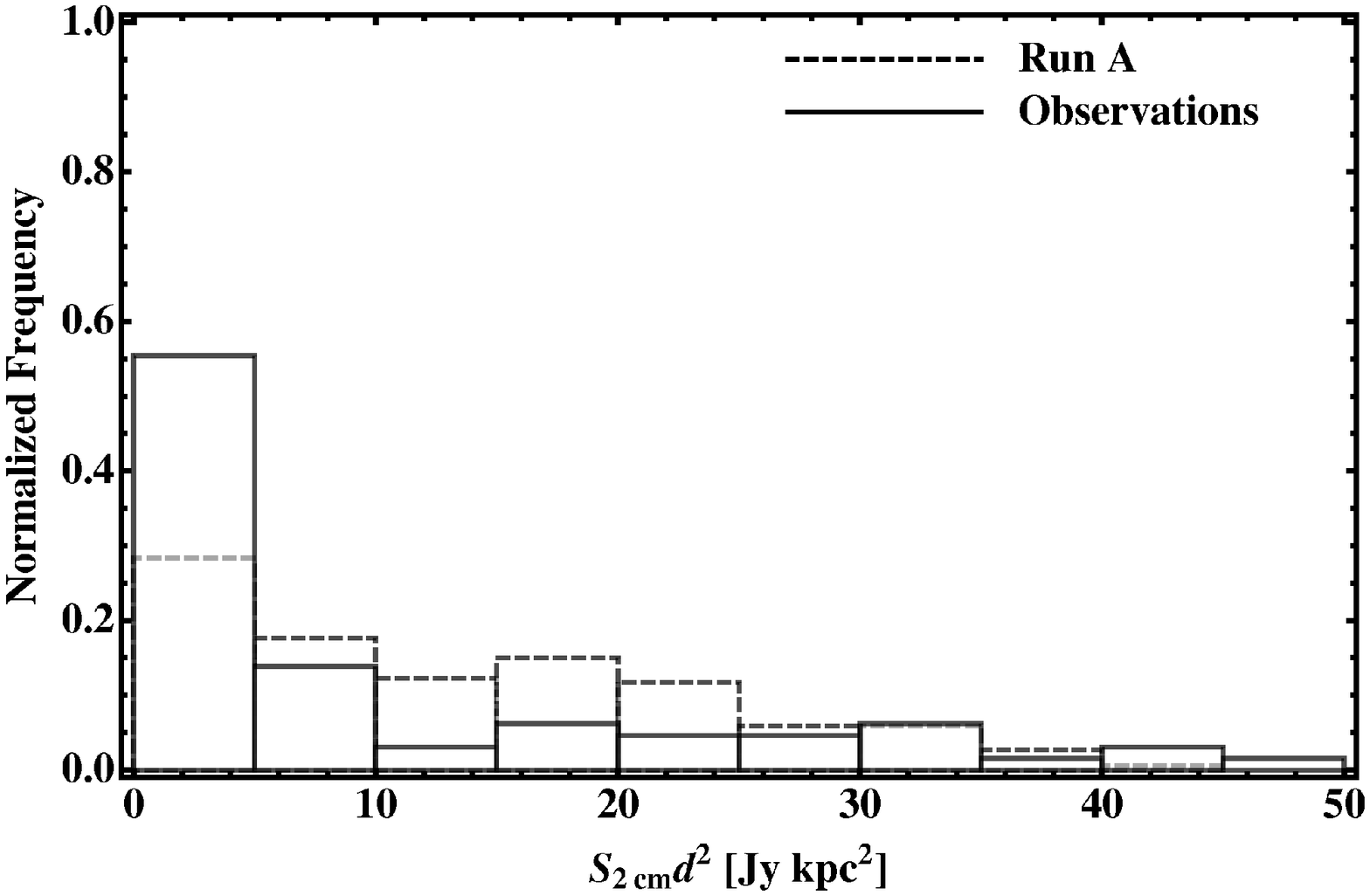}
\includegraphics[width=84mm]{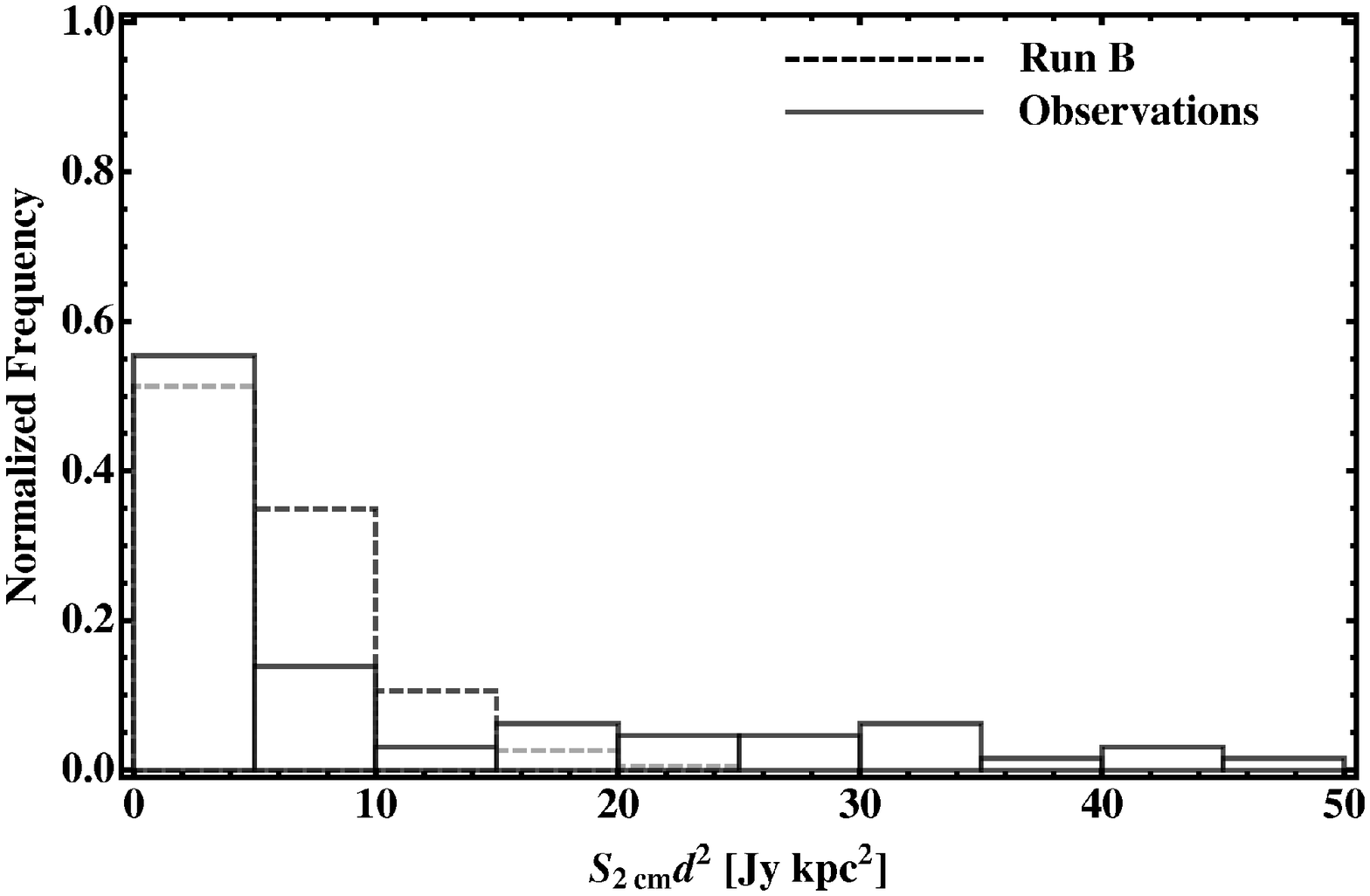}
 \caption{Histograms of 2-cm luminosities ($S_{\rm 2cm}d^2$) for the global temporal evolution of 
 the simulated \HII regions ({\it dashed} lines) and the co-added samples of Wood \& Churchwell (1989) 
 and Kurtz et al. (1994) ({\it solid} lines). The {\it top} and {\it bottom} frames correspond to Run A 
and Run B respectively.}
  \label{fig5}
\end{figure}

The \HII region is always faint ($S_{\rm 2cm}<1$ Jy at the assumed distance of 
2.65 kpc) for $M_\star<25~\Msun$. Past this point, the \HII region is brighter than 
1 Jy 86 \% of the time (Fig. 1). 

\begin{figure}
\includegraphics[width=84mm]{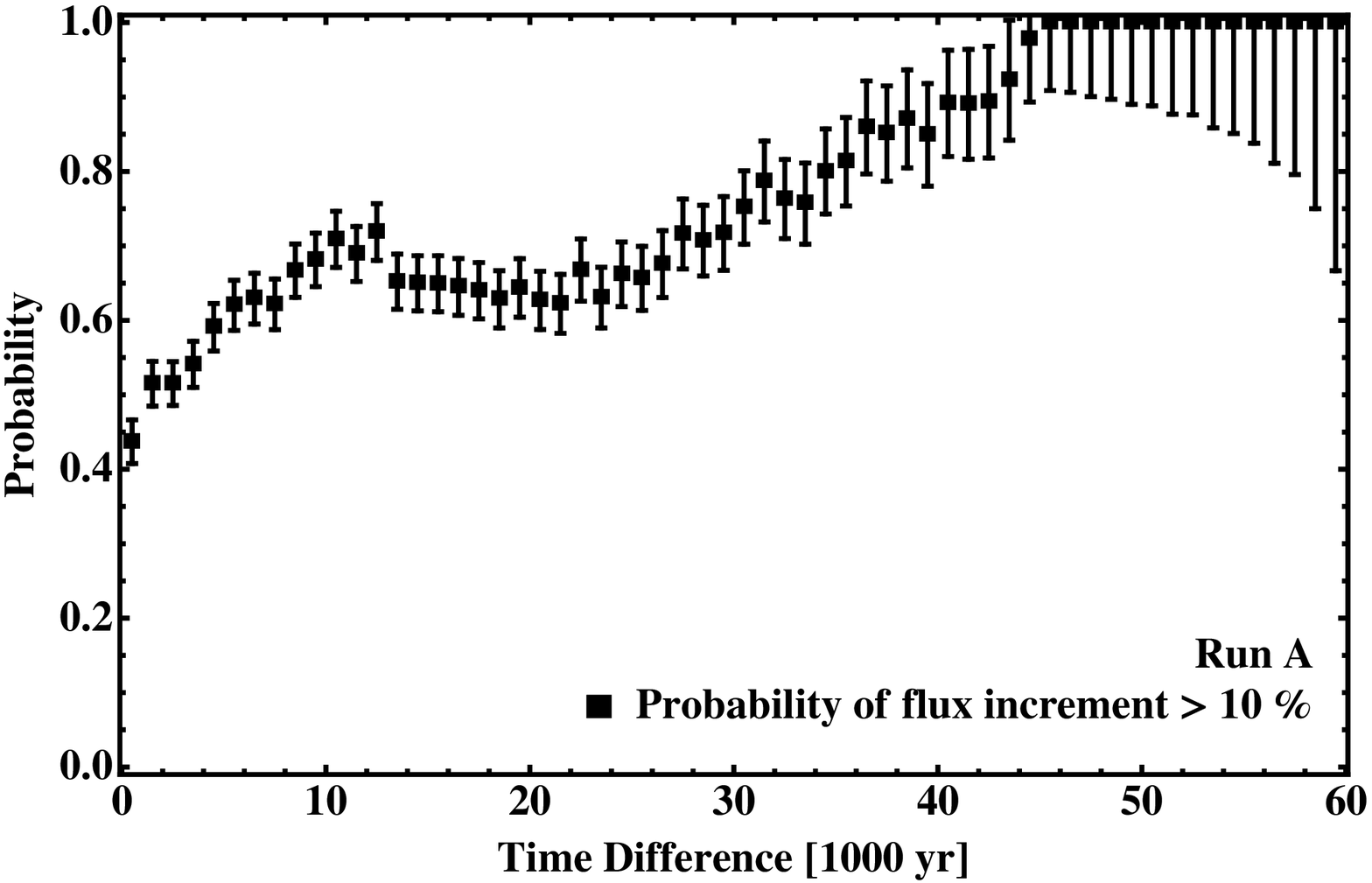}
\includegraphics[width=84mm]{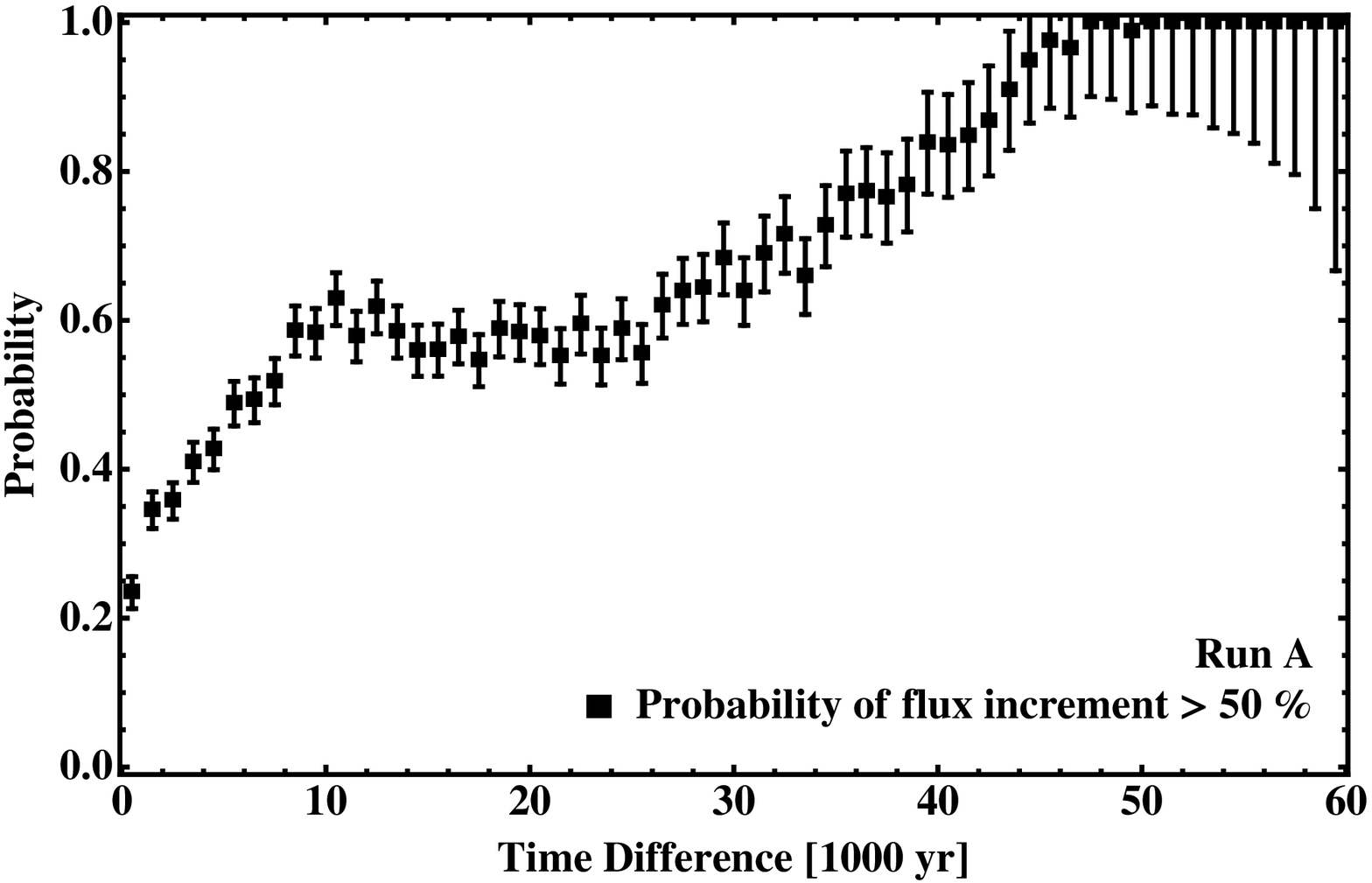}
\includegraphics[width=84mm]{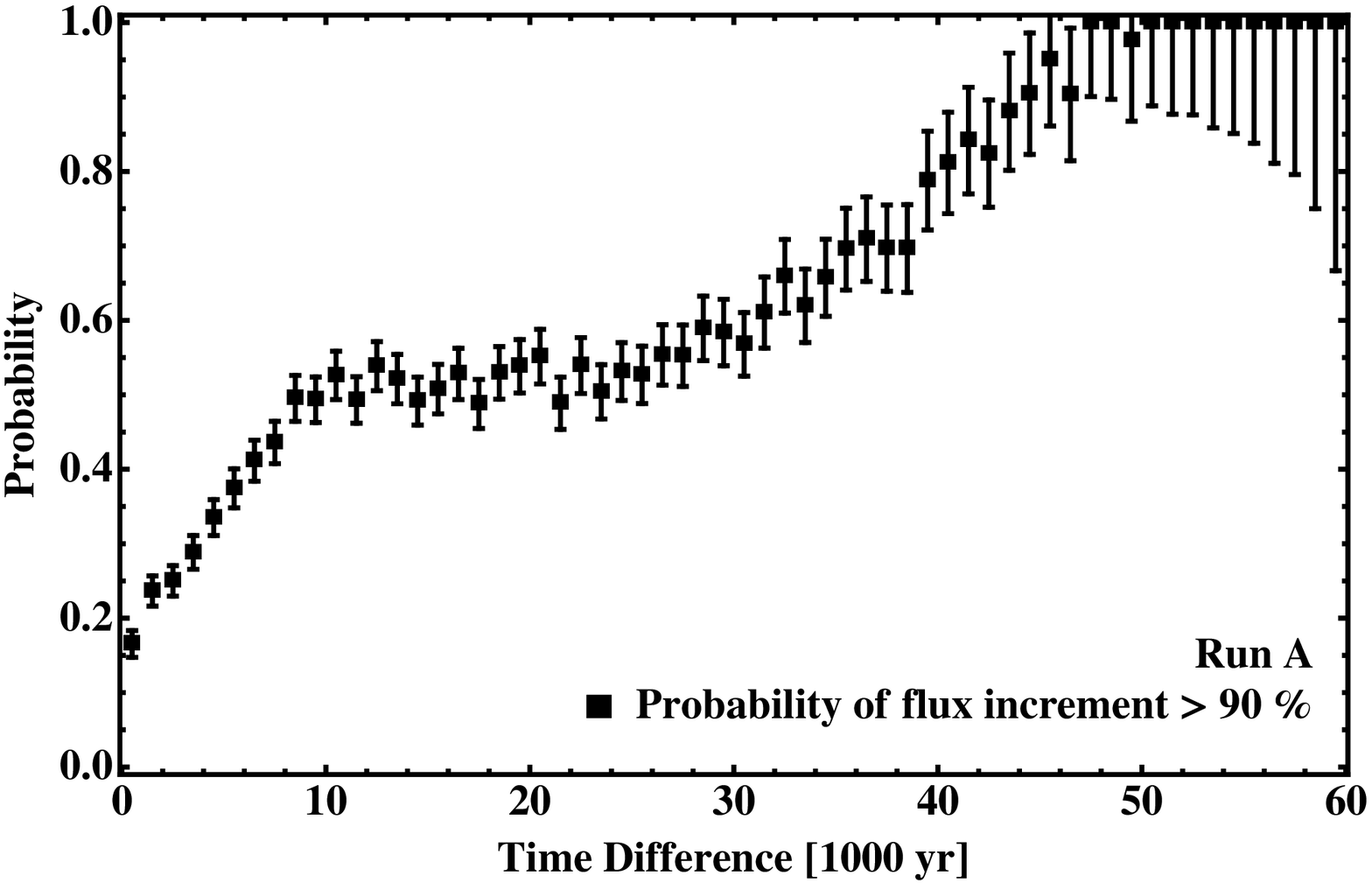}
 \caption{Probabilities for flux increments larger than 10 (top panel), 50 (middle), and $90~\%$ (bottom) as a function 
of time difference for the long-term evolution of the \HII region in Run A. The error bars indicate the $1\sigma$ 
statistical uncertainty from the number of counts in each bin 1000-yr wide.}
  \label{fig6}
\end{figure}

\begin{figure}
\includegraphics[width=84mm]{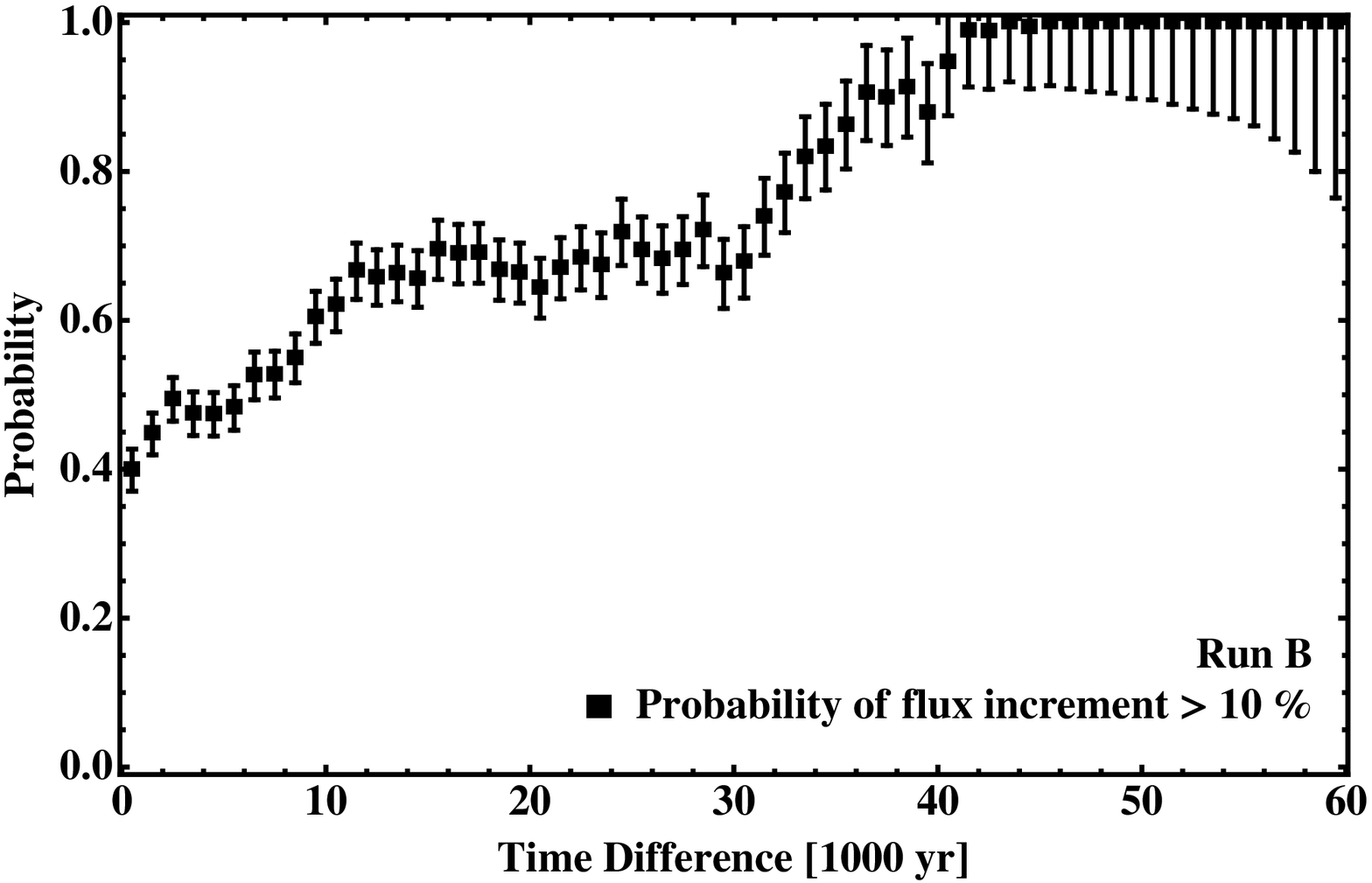}
\includegraphics[width=84mm]{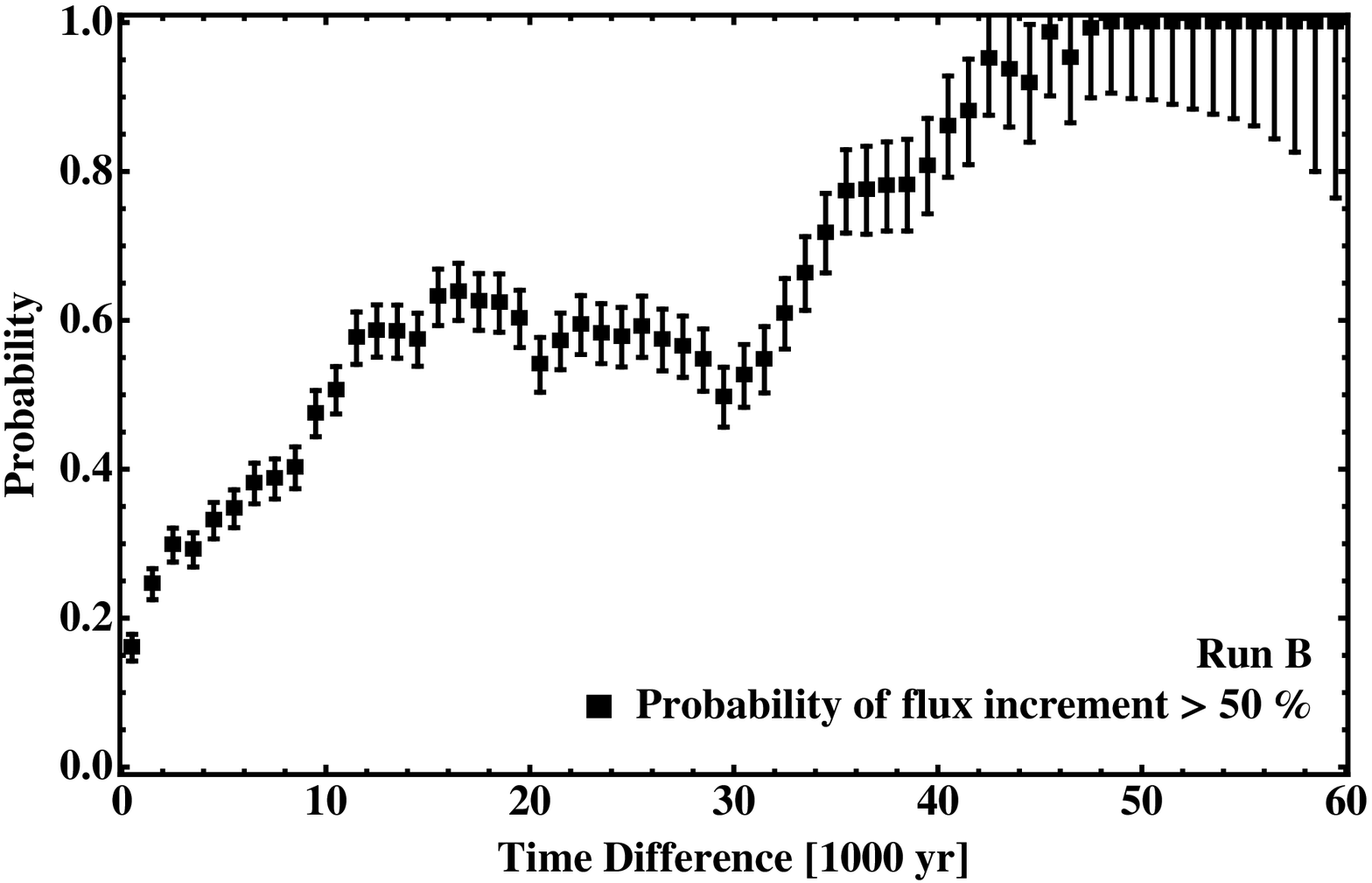}
\includegraphics[width=84mm]{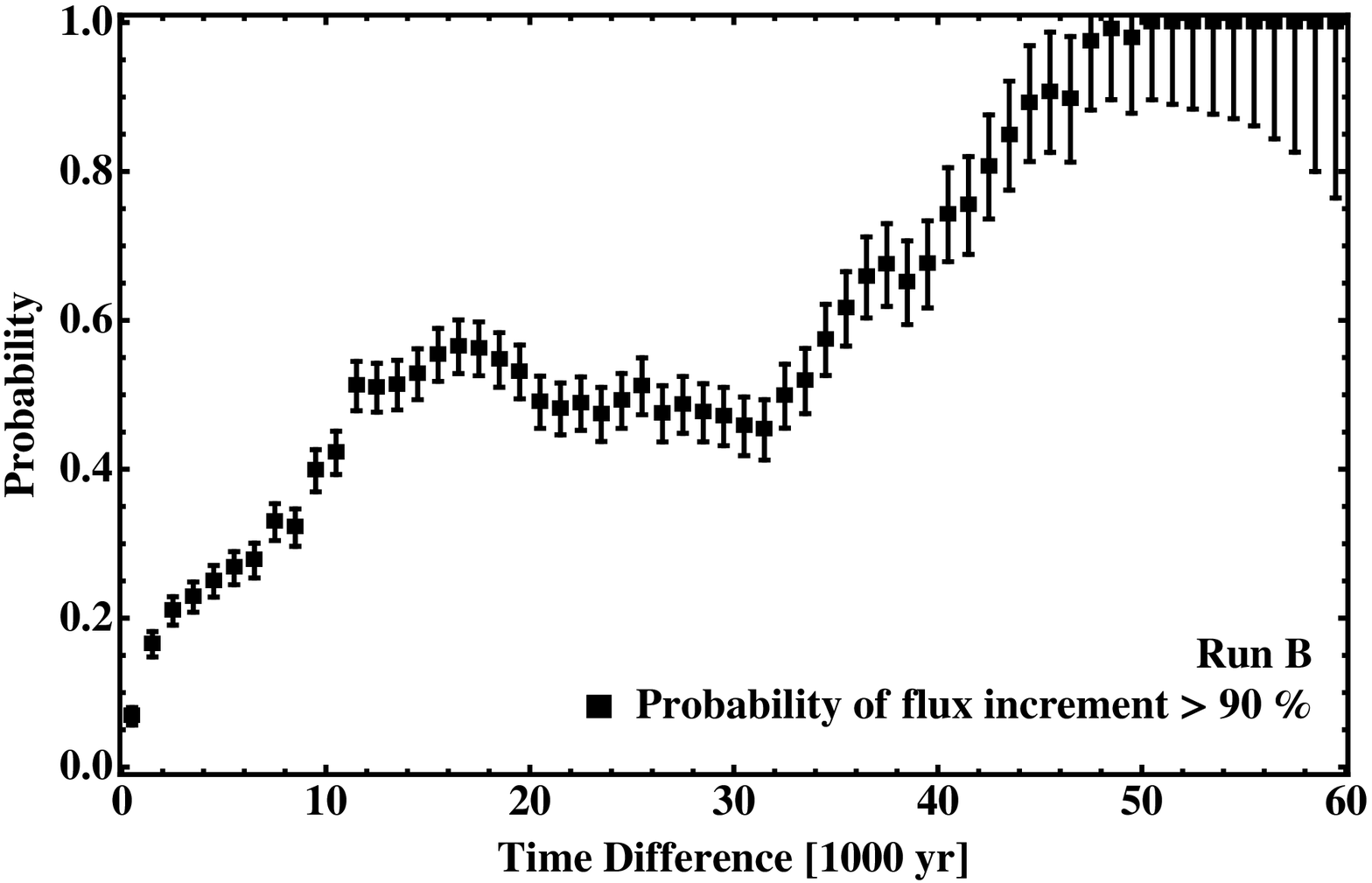}
 \caption{Probabilities for flux increments larger than 10 (top panel), 50 (middle), and $90~\%$ (bottom) as a function 
of time difference for the long-term evolution of the \HII regions in Run B. The error bars indicate the $1\sigma$ 
statistical uncertainty from the number of counts in each bin 1000-yr wide.}
  \label{fig7}
\end{figure}

\begin{figure}
\includegraphics[width=84mm]{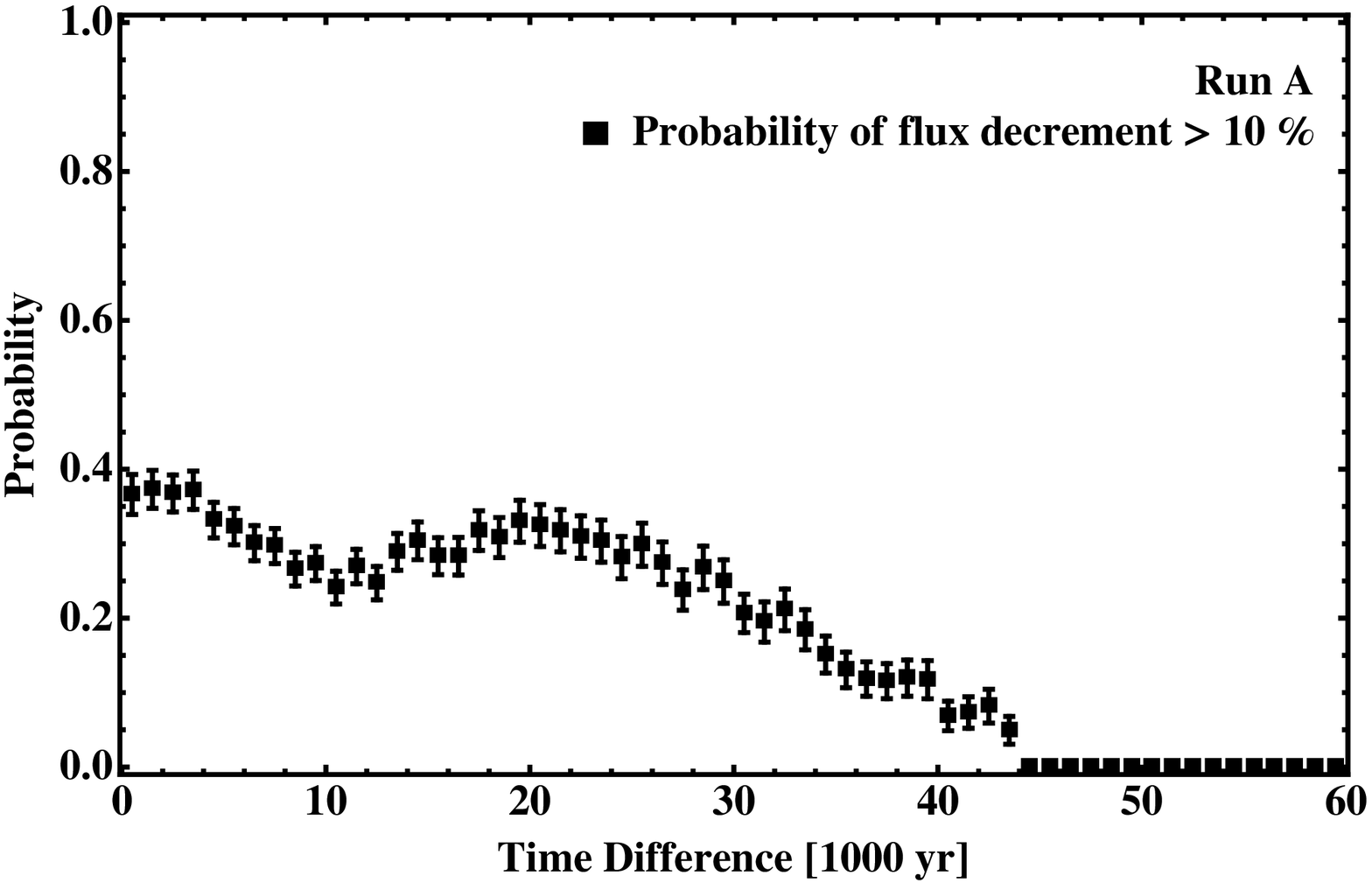}
\includegraphics[width=84mm]{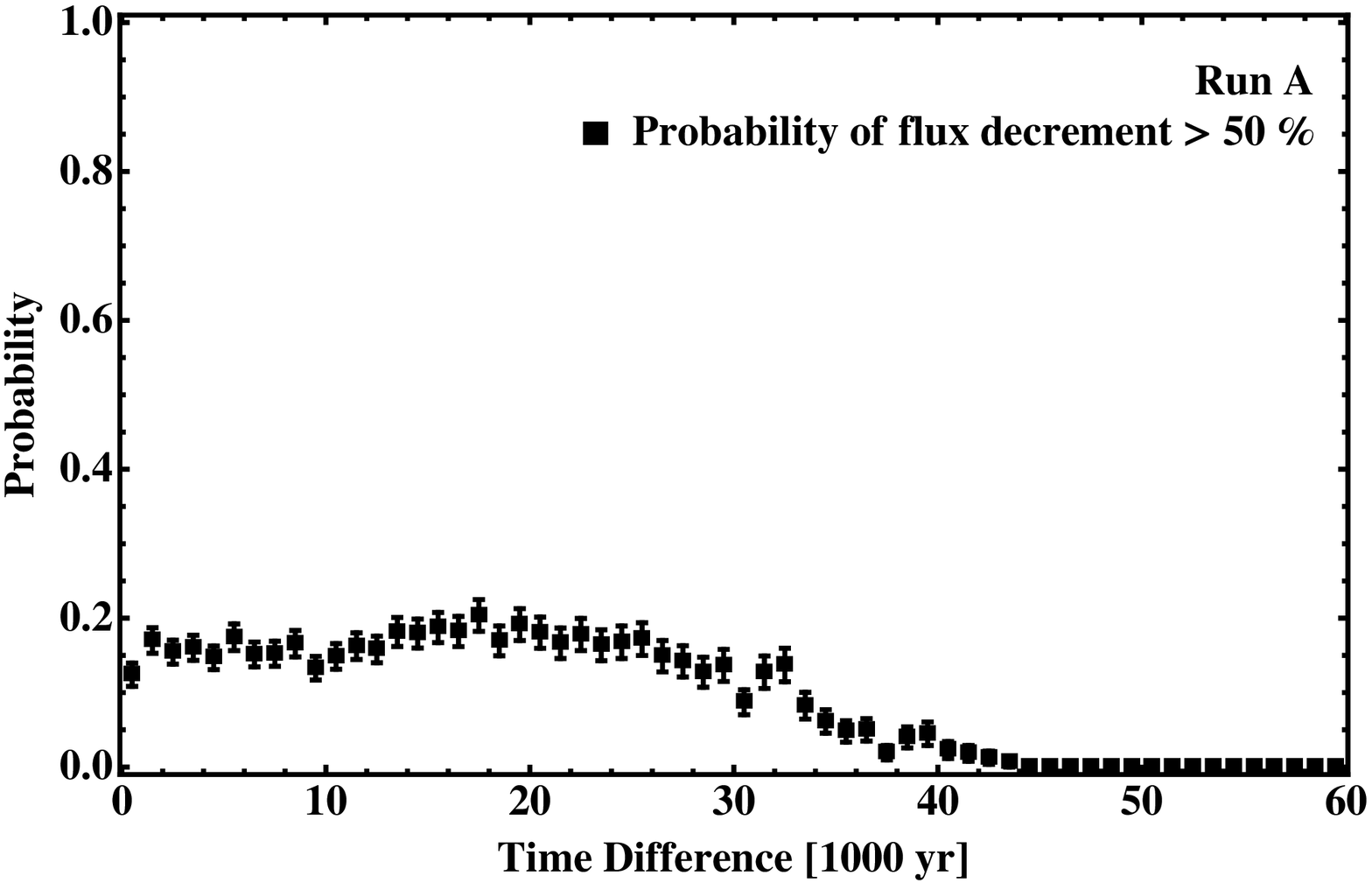}
\includegraphics[width=84mm]{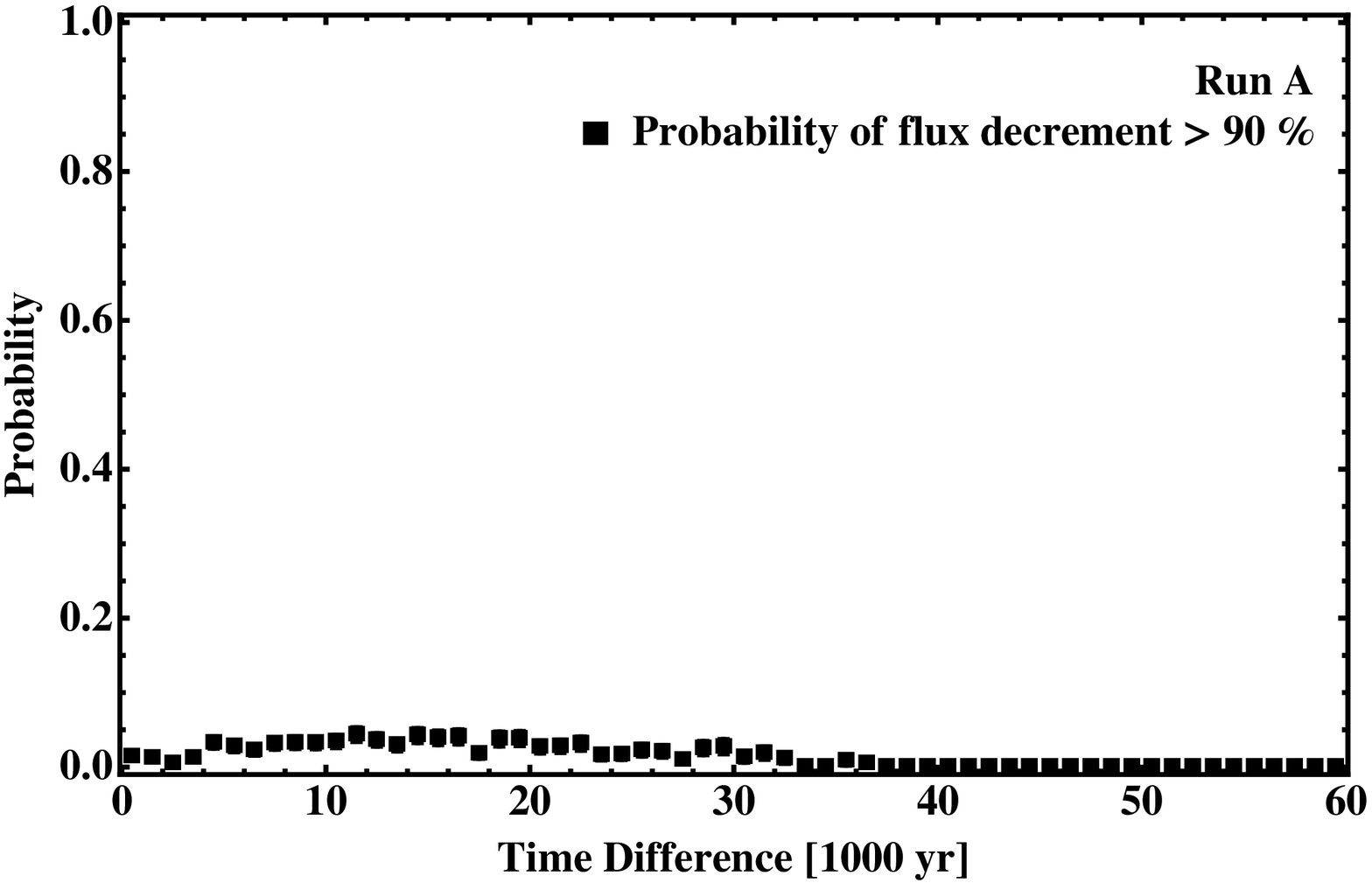}
 \caption{Probabilities for flux decrements larger than 10 (top panel), 50 (middle), and $90~\%$ (bottom) as a function 
of time difference for the long-term evolution of the \HII region in Run A. The error bars indicate the $1\sigma$ 
statistical uncertainty from the number of counts in each bin 1000-yr wide.}
  \label{fig8}
\end{figure}

\begin{figure}
\includegraphics[width=84mm]{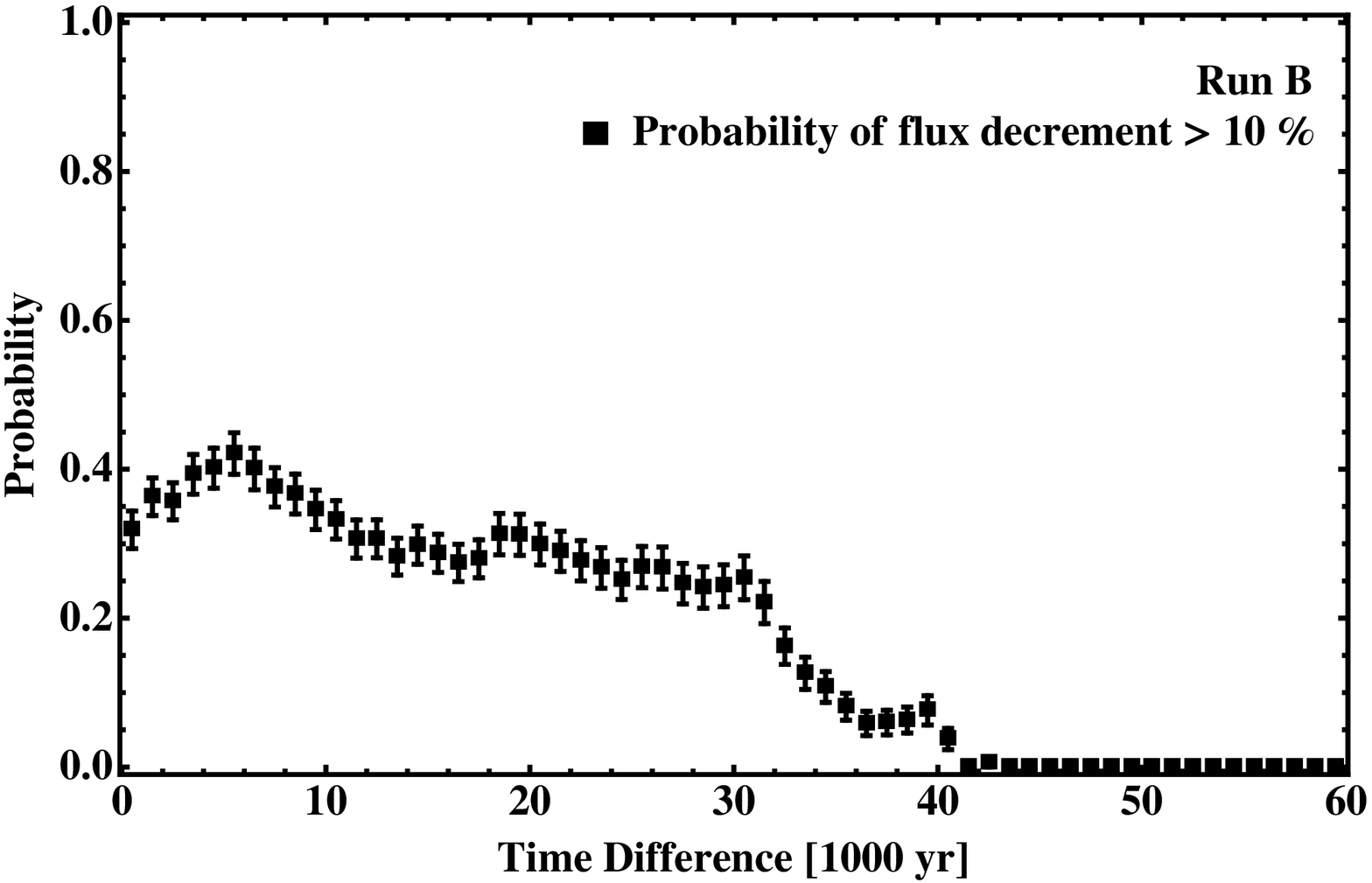}
\includegraphics[width=84mm]{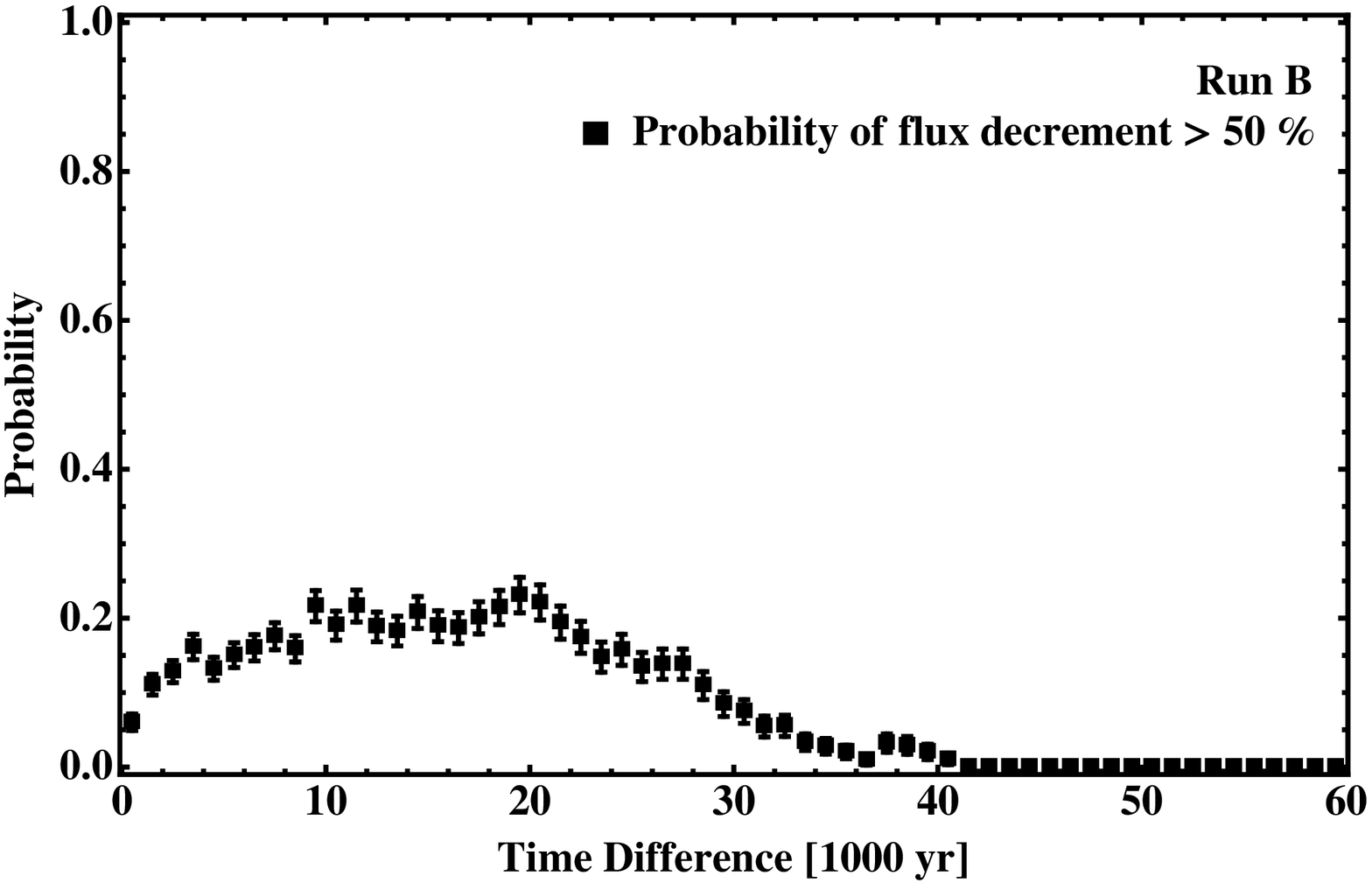}
\includegraphics[width=84mm]{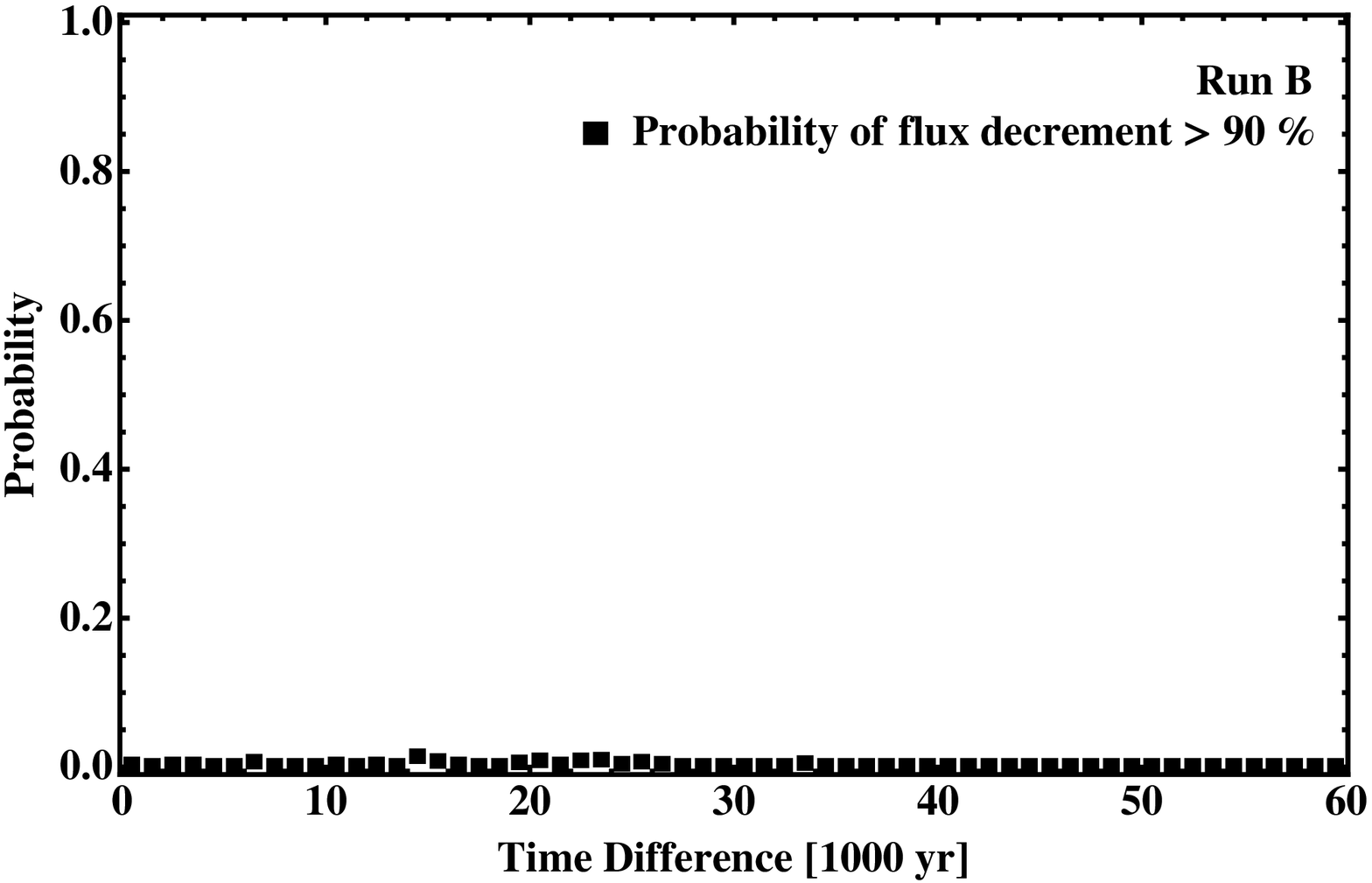}
 \caption{Probabilities for flux decrements larger than 10 (top panel), 50 (middle), and $90~\%$ (bottom) as a function 
of time difference for the long-term evolution of the \HII regions in Run B. The error bars indicate the $1\sigma$ 
statistical uncertainty from the number of counts in each bin 1000-yr wide.}
  \label{fig9}
\end{figure}

A similar analysis of the \HII region around the most massive star in Run B (multiple sinks) 
is shown in Fig. 2. Only the Z-axis projection, i.e., a line of sight perpendicular 
to plane of the accretion flow is used, since only from this viewing angle the brightest \HII 
region is well separated from other \HII regions at all times (the flux 
movies are presented in Paper I). 
The extra fragmentation in Run B translates into a weaker accretion flow and a lower-mass 
ionizing star as compared to that of Run A, a process referred to as 'fragmentation-induced 
starvation' (a theoretical discussion of this process in presented in Paper III). Therefore, 
the brightest \HII region in Run B (Fig. 2) is weaker than the \HII region in Run A (Fig. 1). 
Figures 1 and 2 do not show times later than $t=100$ kyr in order to facilitate their 
comparison. Both runs continue past this time, but in Run B accretion onto the most massive star 
stops at $t=109$ kyr, while in Run A the artificial suppression of the fragmentation leads to 
an unrealistically large mass for the ionizing star at later times (see Paper I).

\HII regions are highly variable both in Run A and Run B. However, since the suppression of 
fragmentation in Run A produces a larger accretion flow and a most massive ionizing star, Run 
A presents larger flux variations than Run B. 
Figure 3 shows the flux changes over the evolution of the \HII regions in both runs.
A comparison of the fractional variations of the \HII regions shows that, though more similar 
between runs, they are still larger in Run A (Figure 4). 
For consecutive data points, in Run A positive flux variations are 56 \% of 
the events and the flux increment is  
$+74$ \% on average, while negative changes are 44 \% of the events and have an average magnitude of 
$-27$ \%. Similarly, for Run B, positive changes (52 \% of the events) have an average magnitude of 
$+42$ \%, while the average flux decrement (48 \% of the events) is $-18$ \%.

\subsection{Comparison to Surveys}

\indent

We present a comparison with the ionized-gas surveys 
of UC and HC \HII regions by \cite{WC89} and \cite{Kurtz94}\footnote{ 
The relation of the surrounding molecular gas to the ionized gas 
has not been explored in detail for many of their sources, which makes difficult to 
assess whether they are candidates to harbor accreting protostars.}. 
Figure 5 shows normalized histograms of 2-cm luminosity ($S_{\rm 2cm}d^2$) obtained for both runs using the 
time steps previously shown in figures 1 to 4 and the 
co-added observed samples from \cite{WC89} and \cite{Kurtz94}, 
taking  the 81 sources for which 
they report a 2-cm flux and a distance. Simulation and observations roughly agree, but 
neither Run A nor B can reproduce the high-luminosity end of the observed distribution: 20 
\% of the observed sources have $S_{\rm 2cm}d^2>50$ Jy kpc$^2$, while only 1 \% of the \HII regions in the 
simulation steps of Run A and in no step in Run B have luminosities above this threshold. These bright UC \HII regions 
likely correspond to stages in which accretion to the ionizing protostar(s) is completely shut off, therefore 
they do not correspond to the analyzed \HII regions from the simulation in which the protostars are still 
accreting.  
In the range $20<S_{\rm 2cm}d^2<50$ Jy kpc$^2$, Run A matches better than 
Run B the observed luminosities. This may be because Run B does not produce any star with a mass 
$M_\star>30~\Msun$.  Simulations identical to Run B produce higher-mass stars
in the presence of magnetic fields \citep{Peters11} 
and may also produce more massive stars in the purely
radiation-hydrodynamical case with higher-mass initial clumps. 
We plan to perform in the future a study on the robustness 
of our results for different initial conditions. 
For the smallest luminosities, Run B matches better than Run A the observed 55 \% of \HII regions that have  
$S_{\rm 2cm}d^2<5$ Jy kpc$^2$, but over-estimates (35 \%) the observed fraction (14 \%) of \HII regions 
with $5<S_{\rm 2cm}d^2<10$. 
\cite{Peters10b} presented statistics of the morphologies of the \HII regions in Run A and Run B and found 
that Run B agrees better with observed surveys. Although Run A can be interpreted as a mode of 
isolated massive star formation, the treatment of fragmentation is more realistic in Run B (see Paper I). 
Moreover, most high-mass stars form in clusters \citep{ZY07}.

\subsection{Long-Term Variation Probabilities}
\indent

We address the question of the flux variability expected from the simulations by calculating  
the probability of variations larger than a given threshold as a function of 
time difference between steps. 
The low time-resolution data has the advantage of spanning the entire runs, but is not 
useful to predict the expected variations on timescales shorter than $10^2$ yr. We use 
the high time-resolution data sets to make an estimate of the flux variations over shorter timescales, 
but we caution that the analyzed time intervals may not be representative of the entire simulation. 
We use this approach 
because re-running the entire simulations to produce data at $\sim 10$ yr resolution 
is not feasible.

Figures 6 and 7 show the probabilities of {\it flux increments larger than a given threshold} for time differences 
between 1 and 60 kyr for Run A and Run B respectively. The {\it top} panels correspond to flux increments 
larger than $10~\%$, the {\it middle} panels correspond to $50~\%$, and the {\it bottom} panels to 
$90~\%$. On average, \HII regions tend to expand, making a given flux increment to be more likely to happen 
for larger time intervals than for shorter ones.

Figures 8 and 9 show the probabilities of {\it flux decrements larger than a given threshold} 
(in modulus) 
for Run A and Run B respectively.  
At $\Delta t > 30$ kyr the negative-change probabilities decrease and 
reach $\sim 0$ at about $\Delta t > 40$ kyr for any threshold. This is caused by the eventual 
growth of the \HII regions in spite of the flickering. 
There is some indication that the
probabilities for negative changes also decrease at timescales shorter than 1 kyr, 
specially for flux-change thresholds larger than $50~\%$ (see Figs. 
8 and 9). This is also suggested from the analysis of the high temporal-resolution data in the next section. 
Although flux increments are more likely than decrements for any given threshold and 
time lag, a novel result of these simulations is that negative flux changes do happen, in contrast 
with the simple expectation of ever growing \HII regions.

\subsection{Short-Term Variation Probabilities}

\begin{figure}
\includegraphics[width=84mm]{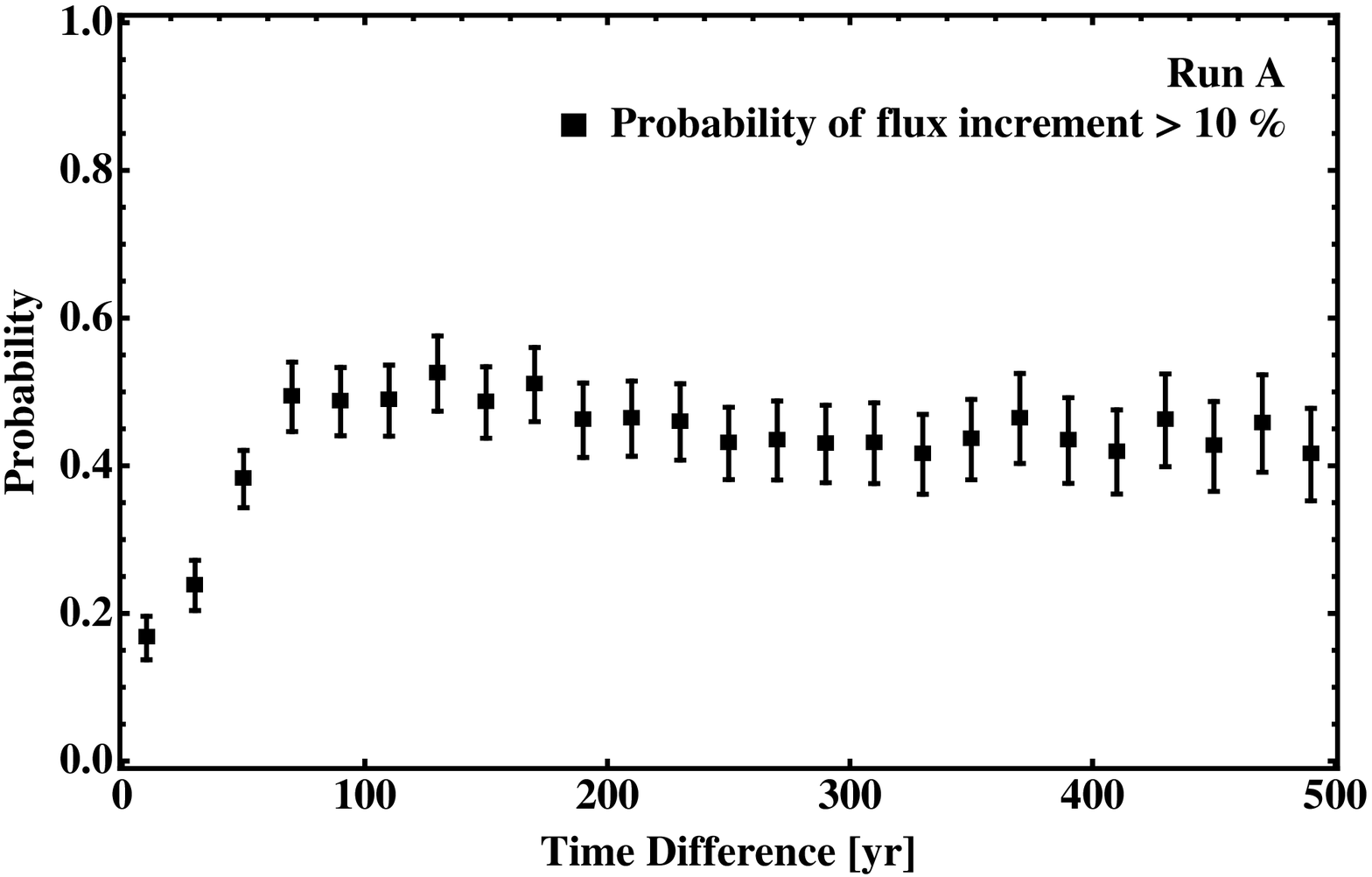}
\includegraphics[width=84mm]{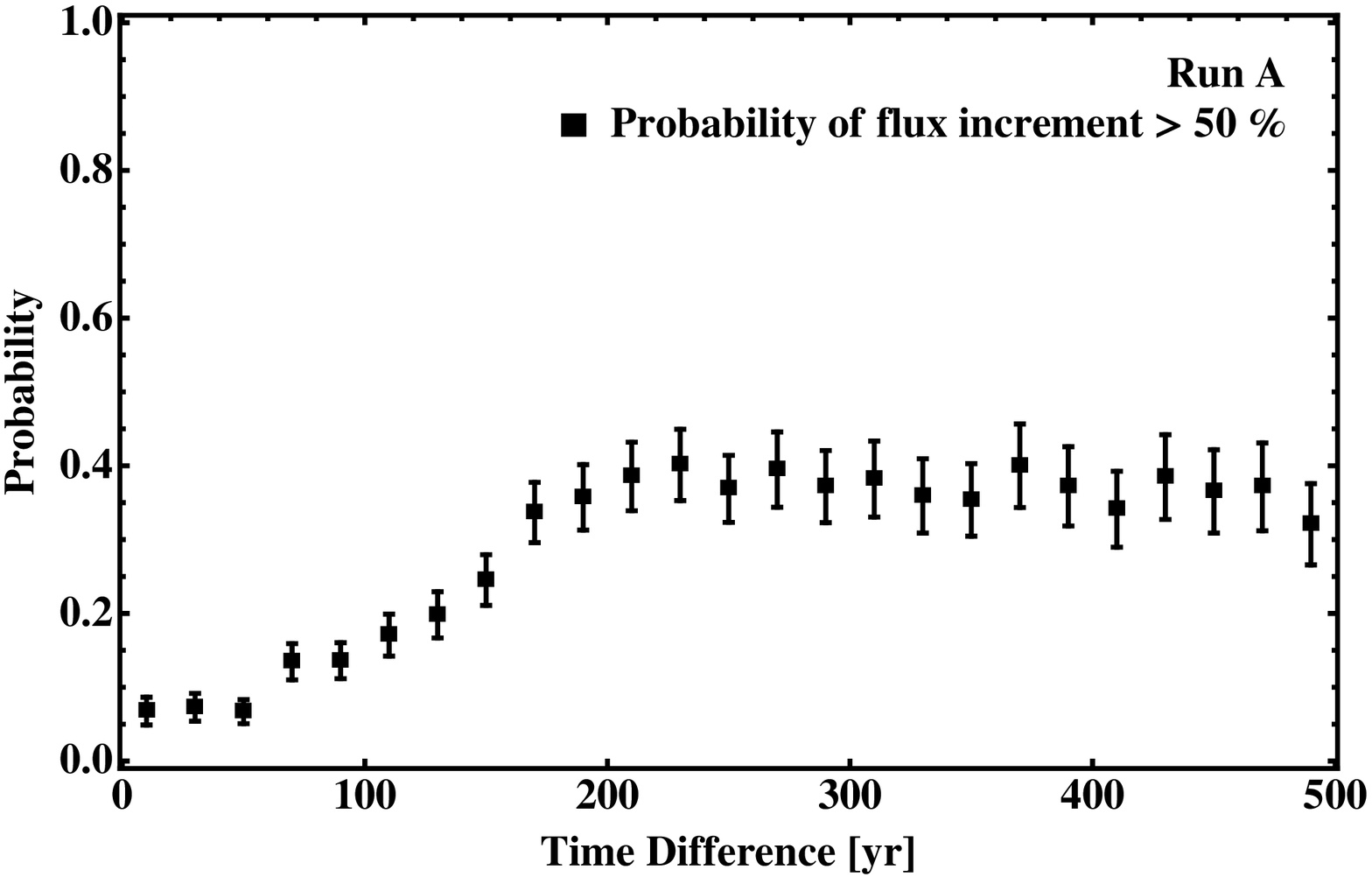}
\includegraphics[width=84mm]{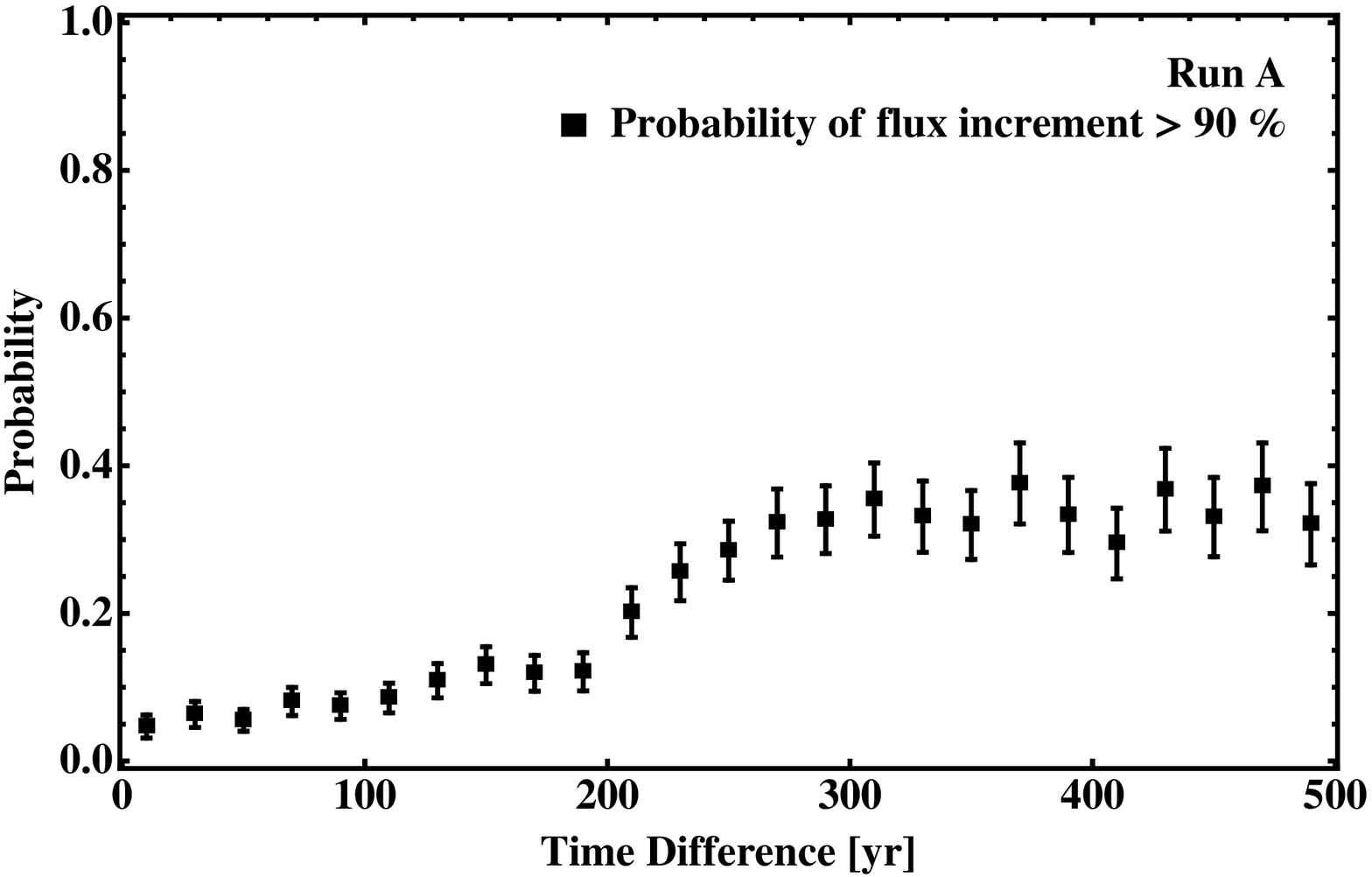}
 \caption{Probabilities for flux increments larger than 10 (top panel), 50 (middle), and $90~\%$ (bottom) as a function 
of time difference for the sample intervals at high time resolution.  The error bars indicate the $1\sigma$ 
statistical uncertainty from the number of counts in each bin 20-yr wide.}
  \label{fig10}
\end{figure}

\begin{figure}
\includegraphics[width=84mm]{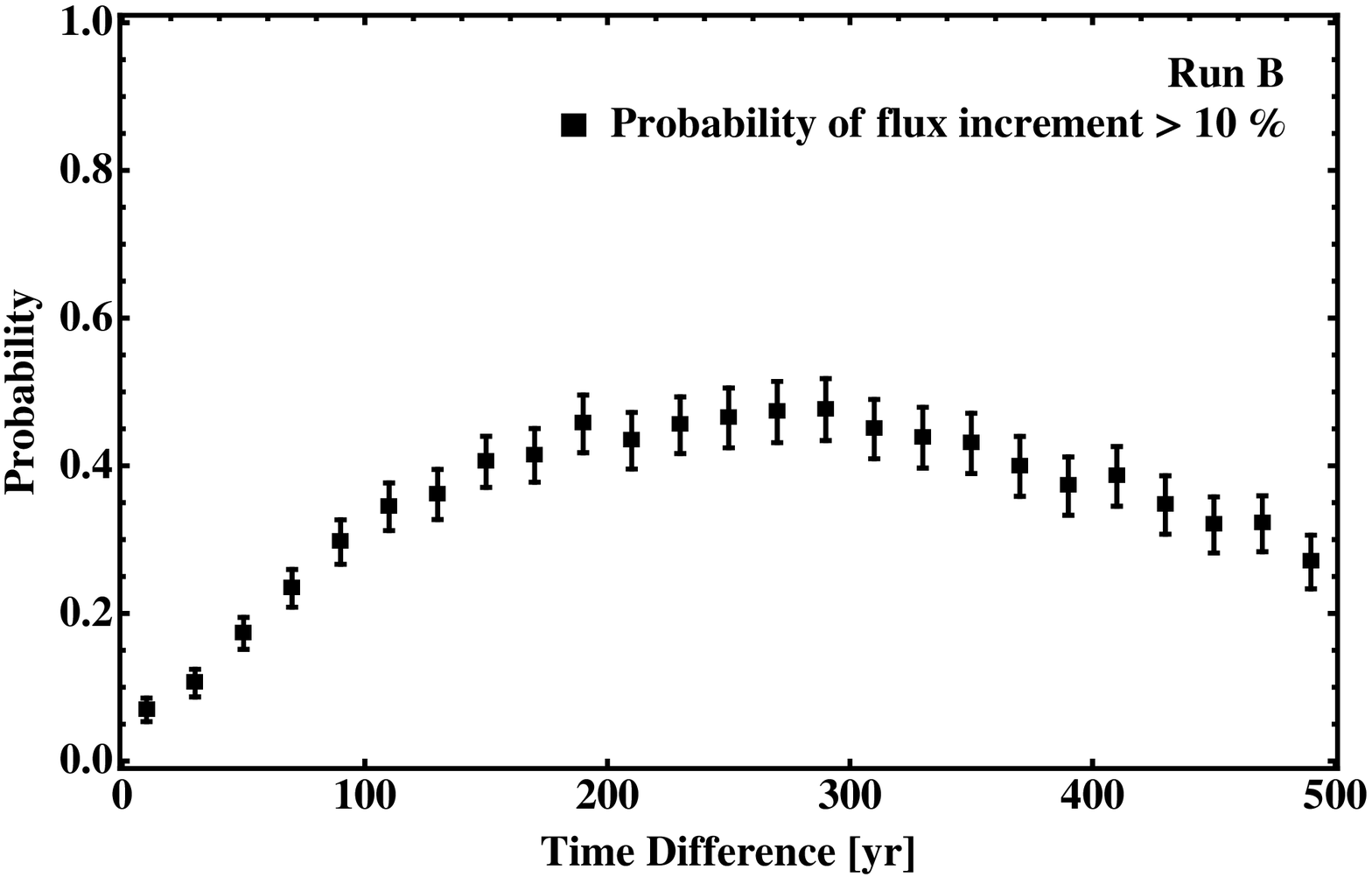}
\includegraphics[width=84mm]{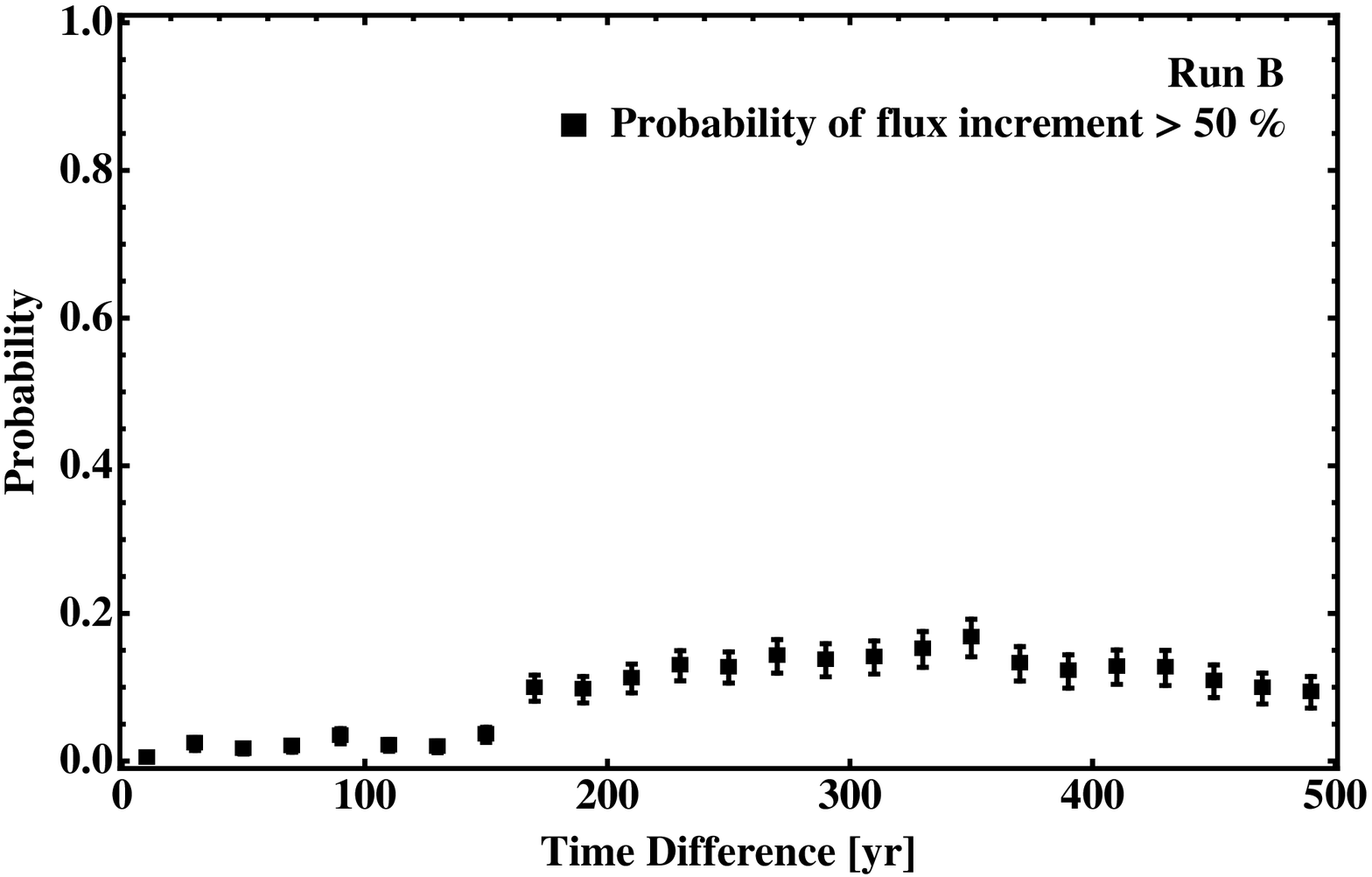}
\includegraphics[width=84mm]{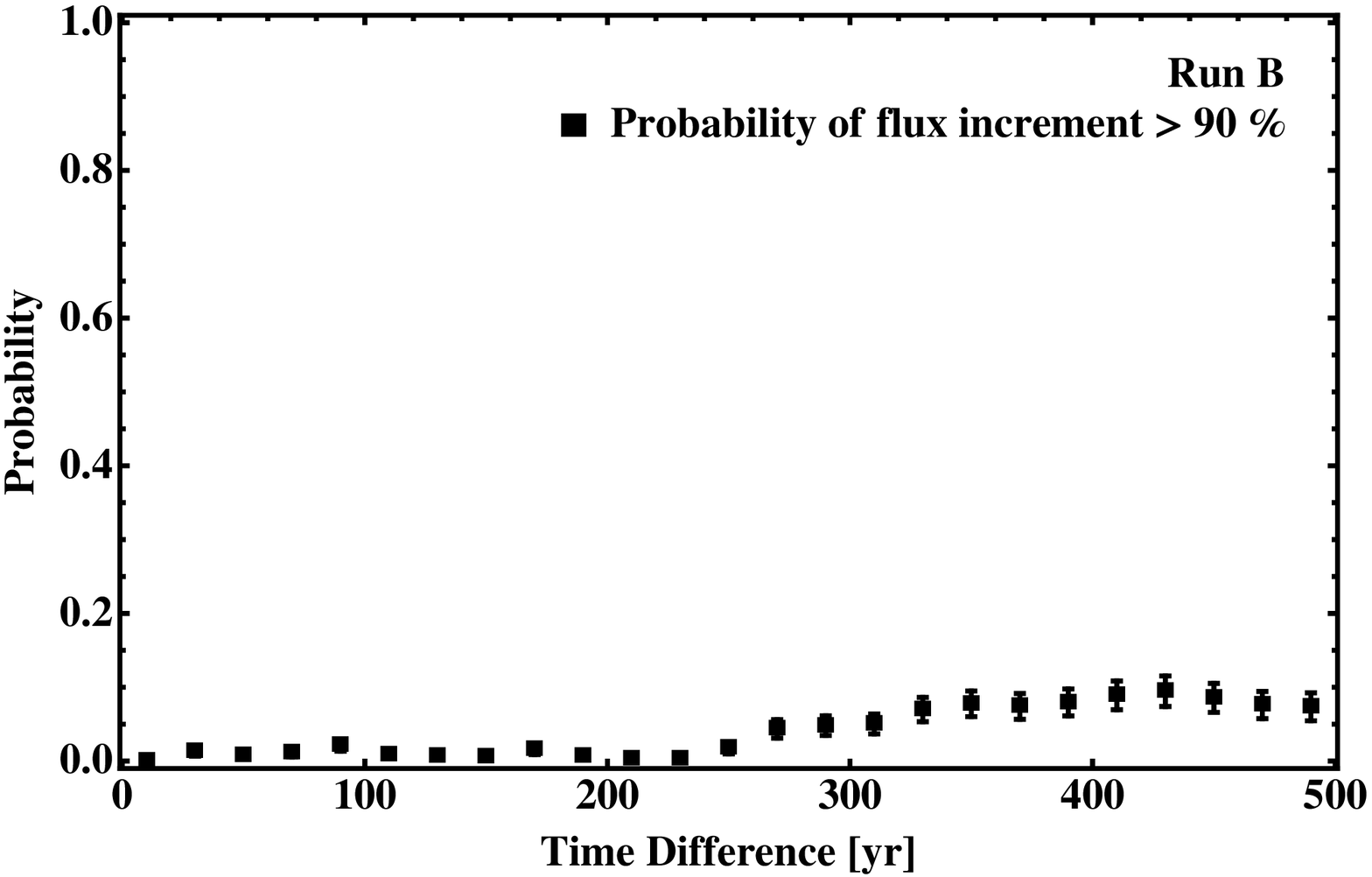}
 \caption{Probabilities for flux increments larger than 10 (top panel), 50 (middle), and $90~\%$ (bottom) as a function 
of time difference for the sample intervals at high time resolution, for Run B.  The error bars indicate the $1\sigma$ 
statistical uncertainty from the number of counts in each bin 20-yr wide.}
  \label{fig11}
\end{figure}

\begin{figure}
\includegraphics[width=84mm]{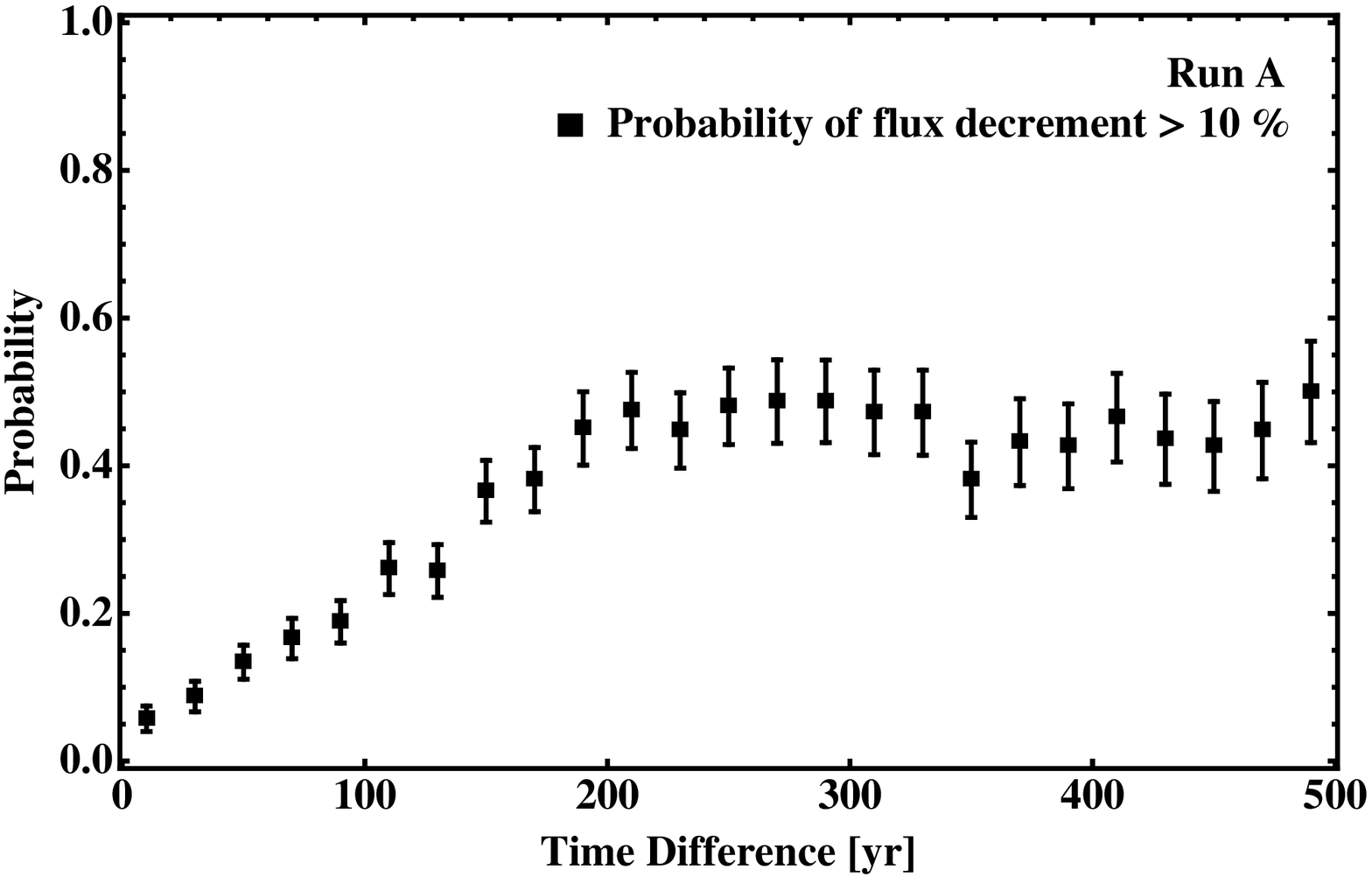}
\includegraphics[width=84mm]{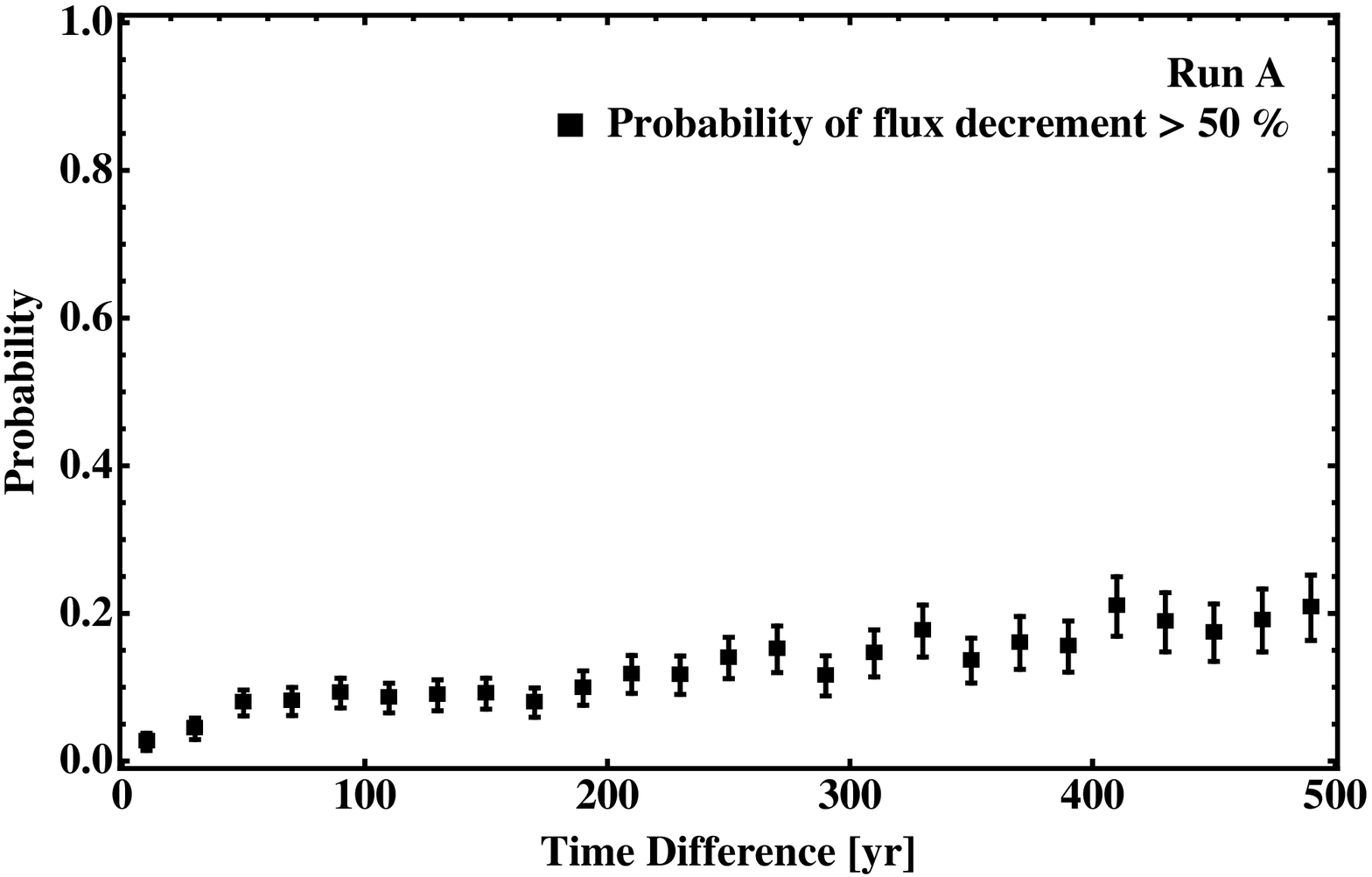}
\includegraphics[width=84mm]{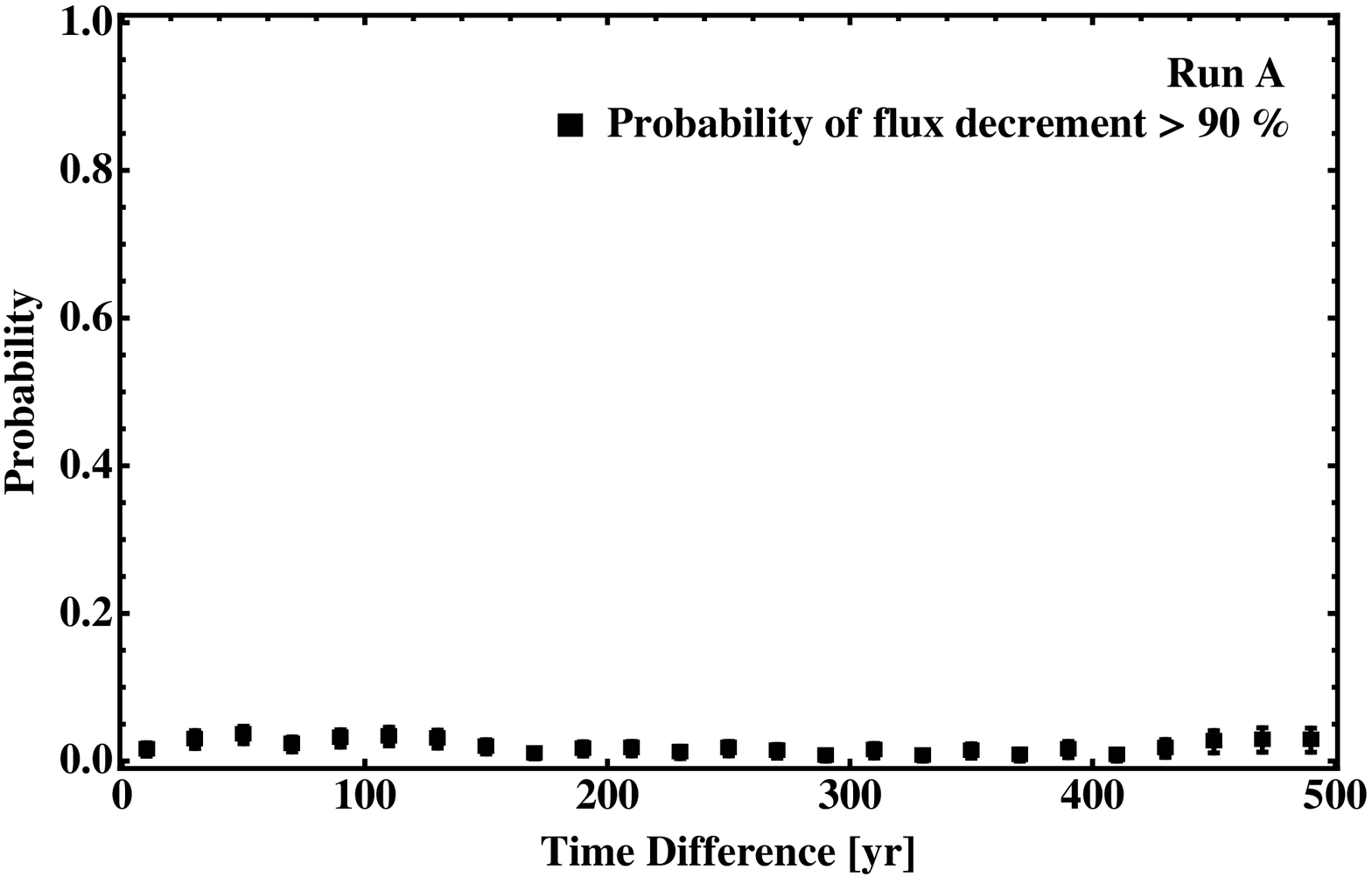}
 \caption{Probabilities for flux decrements larger than 10 (top panel), 50 (middle), and $90~\%$ (bottom) as a function 
of time difference for the sample intervals at high time resolution.  The error bars indicate the $1\sigma$ 
statistical uncertainty from the number of counts in each bin 20-yr wide.}
  \label{fig12}
\end{figure}

\begin{figure}
\includegraphics[width=84mm]{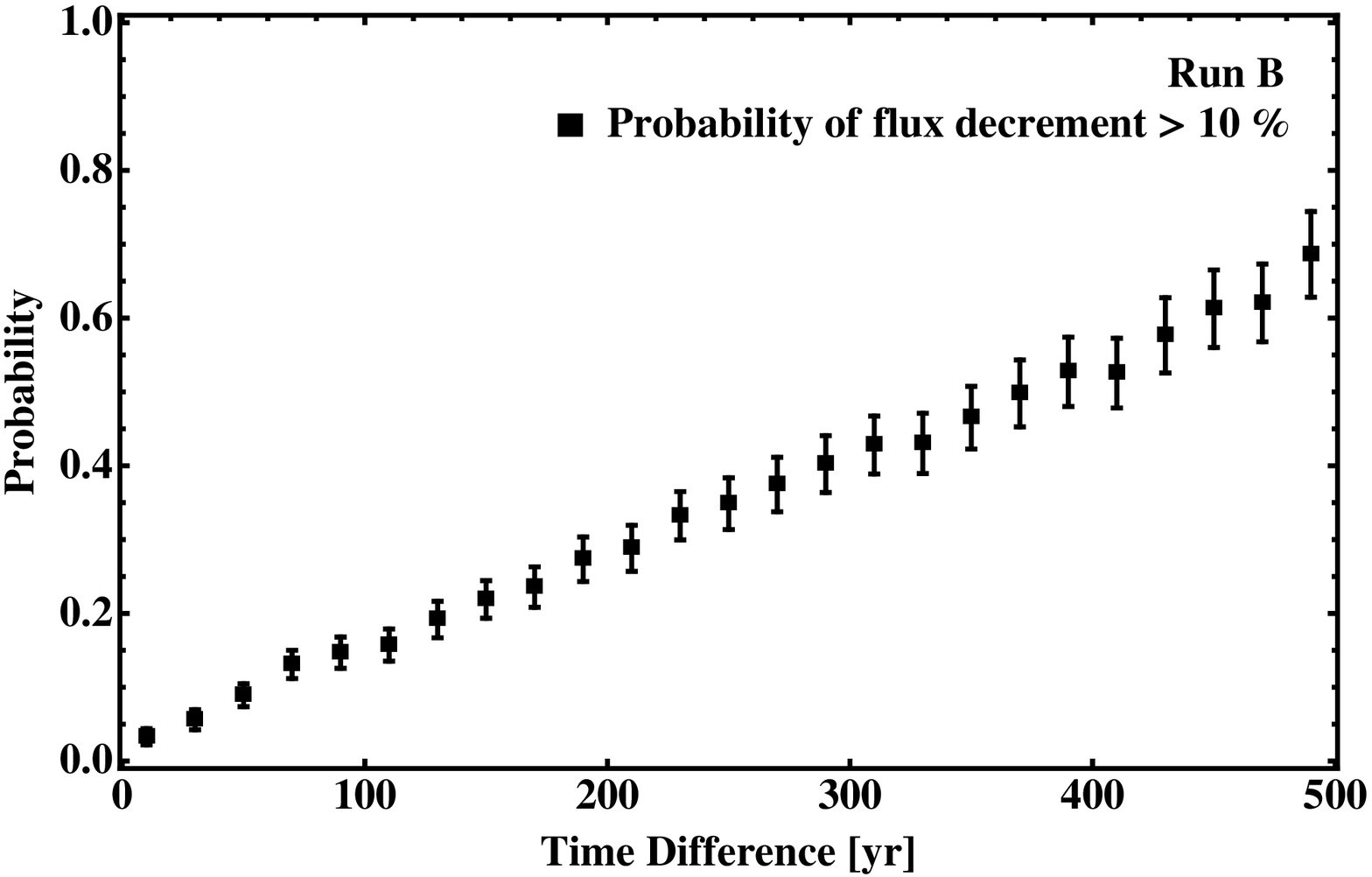}
\includegraphics[width=84mm]{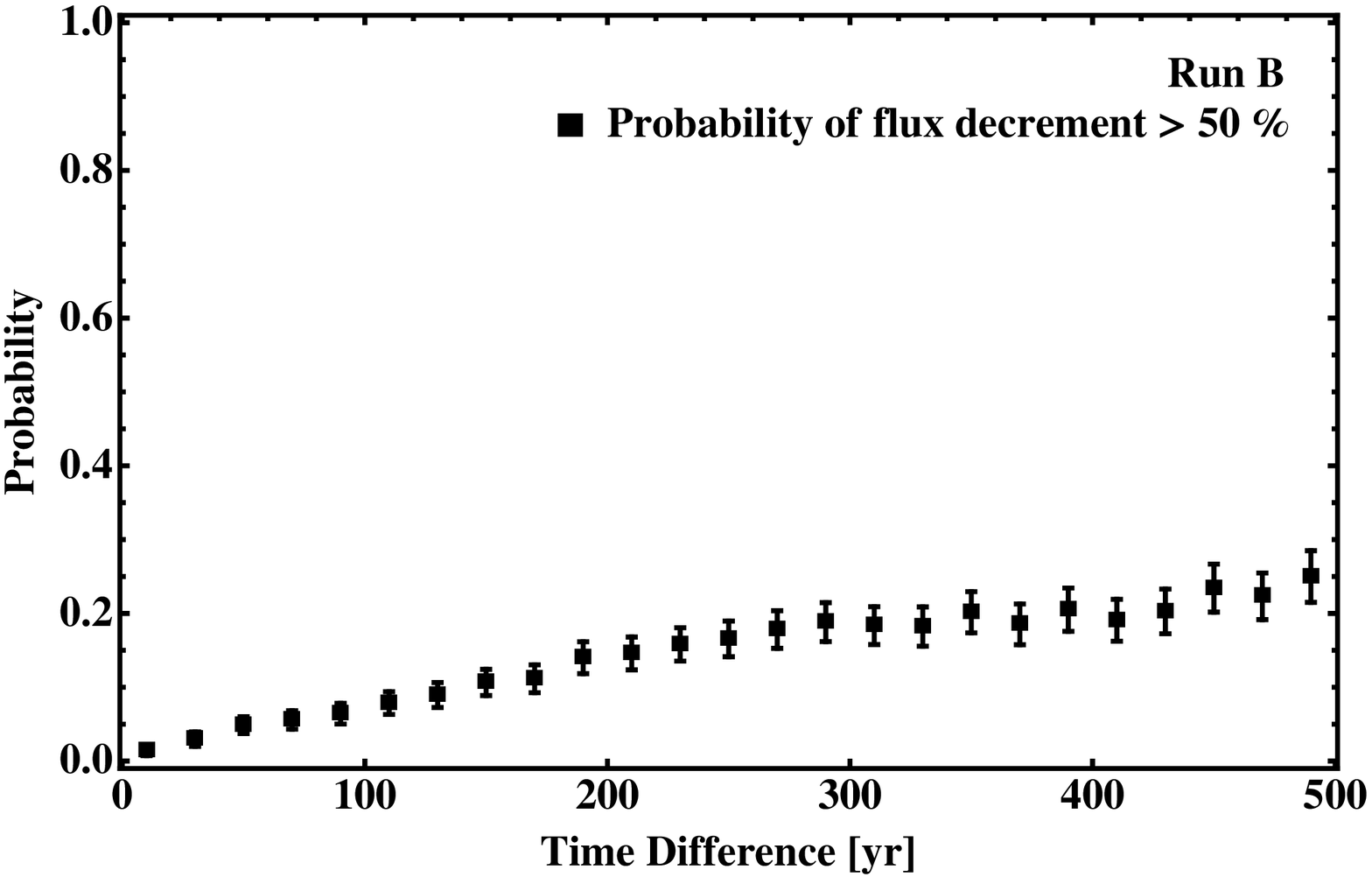}
\includegraphics[width=84mm]{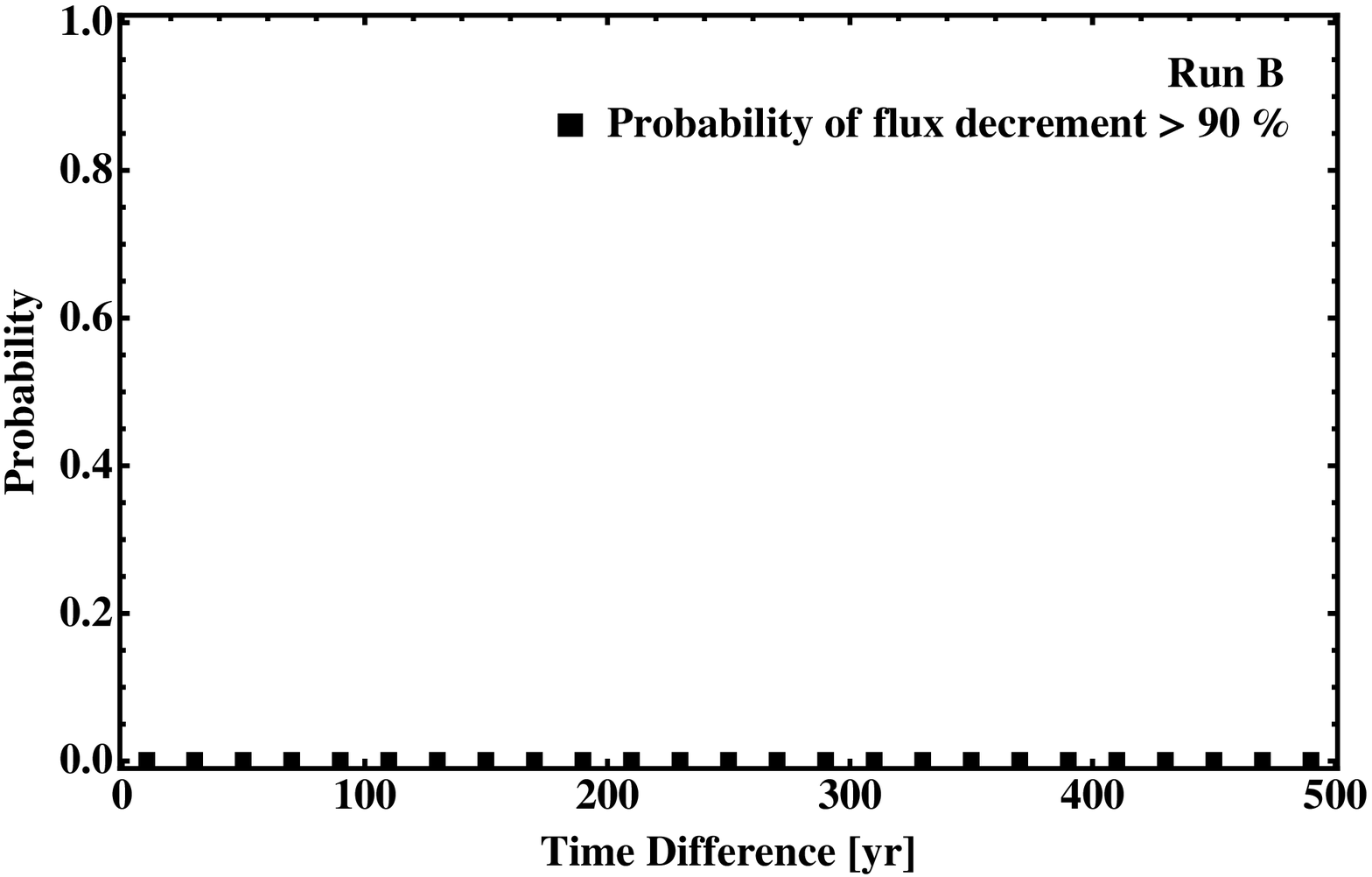}
 \caption{Probabilities for flux decrements larger than 10 (top panel), 50 (middle), and $90~\%$ (bottom) as a function 
of time difference for the sample intervals at high time resolution, for Run B.  The error bars indicate the $1\sigma$ 
statistical uncertainty from the number of counts in each bin 20-yr wide.}
  \label{fig13}
\end{figure}

\indent

We re-ran four time intervals in each of Run A and Run B, spanning a few hundred years each, 
and producing 
data dumps at each simulation step ($\sim 10$ yr). These time 
intervals were selected to contain a pair of negative/positive flux changes to investigate the correlation 
of flux variations with physical changes in the \HII region, like size and density (next section). 
Therefore, they may not be representative of the entire simulation, but since it is not feasible to 
re-run the entire simulations producing data dumps at the highest temporal resolution, we use these  
data to constrain the expected flux variations in observable timescales. We argue 
that the close match at scales of $\Delta t \sim 500$ yr between the probabilities obtained from 
the low temporal-resolution (previous section) and the high temporal-resolution data (this section) 
indicates that the results presented here are meaningful. 

Figures 10 and 11 show, for Run A and Run B respectively, 
the probabilities for flux increments larger than the specified threshold 
as a function of time lag. These figures are the analogs of Figs. 6 and 7 for the 
high temporal-resolution data. The slight probability decrements after $\Delta t \sim 300$ yr, 
especially at low thresholds,  
are an artifact caused by the fact that the data sets include a negative/positive flux-change pair. 
Still, the probabilities at $\Delta t = 490$ yr match within 20 \% to 80 \% with the probabilities 
at $\Delta t = 500$ yr from the low-temporal resolution data.

On observable timescales, $\Delta t = 0$ to 40 yr, a small but non-zero 
fraction of \HII regions is expected to have detectable flux increments. For two observations 
separated by $10$ yr, 
Run A gives a prediction of $16.7\pm2.9~\%$ of \HII regions having flux increments larger than 
10 \%, $6.8\pm1.9~\%$  with flux increments larger than 50 \%, and $4.7\pm1.6~\%$ with increments larger than 
90 \%. 
The more realistic Run B predicts 
a smaller fraction of variable \HII regions:  $6.9\pm1.6~\%$, $0.3\pm0.3~\%$, and $0~\%$ of them are expected to have 
flux increments larger than 10 \%, 50 \%, and 90 \% over a time interval of $10$ yr, respectively.

Figures 12 and 13 show the probabilities for flux decrements larger than a given threshold obtained 
with the high temporal-resolution data. The probabilities obtained at $\Delta t=490$ yr from the high 
temporal-resolution data roughly match with those at $\Delta t=500$ yr obtained from 
the low temporal-resolution data, within a factor of 1 to 3.

Negative variations should also be detectable in a non-negligible fraction of \HII regions. 
Run A predicts that $5.7\pm1.7~\%$, $2.6\pm1.2~\%$, 
and $1.6\pm0.1~\%$ of \HII 
regions should present flux decrements larger than 10 \%, 50 \%, and 90 \% respectively, when observed 
in two epochs separated by $10$ yr. Run B predicts a smaller fraction of 
\HII regions with negative flux variations:  $3.3\pm1.1~\%$, $1.5\pm0.7~\%$, and $0~\%$ for thresholds at 
 10 \%, 50 \%, and 90 \% respectively. We emphasize that negative variations are not expected in 
 classical models of monotonic \HII region growth, 
 while they are a natural outcome of the model presented here. 
 Moreover, these variations should be detectable with current telescopes.

\subsection{Variations in other properties of the HII regions}

\indent 

Because Run B is more realistic in   
the treatment of fragmentation (Section 2.1, see Paper I for details), we  
use the time intervals of Run B with data at high temporal resolution to investigate 
the correlations of sudden flux changes with other properties of the \HII regions.

\smallskip

Let the size of the \HII region of interest be defined as $L_\mathrm{HII}=2(A/\pi)^{1/2}$, 
where $A$ is the area in the synthetic image where the \HII region is brighter than 
three times the rms noise.  Lets also consider the average density of the \HII region 
$\rho_\mathrm{HII}$, and the rate at which neutral gas is converted into ionized gas 
$\dot{M}_\mathrm{\rightarrow HII}$.  Figure 14 compares the time evolution of these 
three quantities.

Figure 15 further compares the time evolution of the 2-cm flux $S_\mathrm{2cm}$, 
the total ionized mass $M_\mathrm{HII}$ in the \HII region, and the denser  
ionized mass within the same volume $M_\mathrm{HII, dense}$. The density 
threshold to define this dense gas is $\rho_\mathrm{HII}>10^{-17}$ g cm$^{-3}$, 
the typical peak density reached when the \HII region gets rapidly 
denser immediately after the flickering events (see Fig. 14).

The flux-size correlation, as well as the size-density anticorrelation are a consequence 
of the relatively large optical depths of the \HII regions (see Section 3.1).
The ``quenching'' events, when an \HII region has a large, sudden drop in flux and size,  
are coincident with large increments in the ionized density. 
The \HII 
region rapidly reaches a state close to ionization equilibrium at its new size after the 
quenching instability. In Fig. 14 it is seen that at the moment of the 
quenching, the ionization rate has a sharp decrement immediately followed by an even faster 
increment that marks the initial re-growth of the \HII region. 
Shortly afterwards, the ionization rate stabilizes again and 
the \HII region grows hydrodynamically, gradually becoming larger and less dense.

The ionized mass of the \HII regions follows the flux and size closely (Fig. 15), 
since this is the gas  
responsible for the free-free emission. However, if only the denser gas is taken into account,  
the amount of this dense ionized gas is particularly high in the initial re-growth of the \HII region 
immediately after the quenching event (Fig. 15). Therefore the rapid quenching events are marked 
by the presence of denser gas around the accreting massive protostar.

\section{Discussion}

\subsection{A new view of early \HII region evolution}

\indent

Until recently, \HII regions have been modeled as (often spherical) bubbles of ionized gas 
freely expanding into a quiescent medium. However, this paradigm fails to explain observations 
of some UC and HC \HII regions (see Section 1). There is enough evidence to assert that massive 
stars form in clusters by accretion of gas from their complex environment 
\citep[reviews have been presented by][]{GL99,MK04,Beuther07,ZY07}. 
Centrally peaked, anisotropic density gradients are expected at the moment an accreting massive star starts 
to ionize its environment. Therefore, 
the earliest \HII regions should not be expected to be the aforementioned bubbles, but more 
complex systems where ionized gas that is outflowing, rotating, or even infalling may 
coexist. The feasibility of this scenario has been shown by analytic models 
\citep{Hollen94,Keto02,Keto03,Keto07,Lugo04} and recent numerical simulations (Papers I, II, and III). 
In these numerical models, the inner part 
of the accretion flow is mostly ionized, while the outer part is mostly neutral. The neutral 
accretion flow continuously tries to feed the central stars. The interaction of this 
infalling neutral gas with the ionized region has a remarkable observational effect: the flickering of 
the free-free emission from the \HII region.  

We stress that the flickering is not a gasdynamical effect. Instead,
it is a non-local result of the shielding of the ionizing source
by its own accretion flow. Hence, variations in the ionization
state of the gas inside the \HII region are not limited by the
speed of sound but can happen on the much shorter recombination
timescale of the ionized gas, rendering 
direct observation of this flickering effect feasible.

\begin{figure}
\includegraphics[width=84mm]{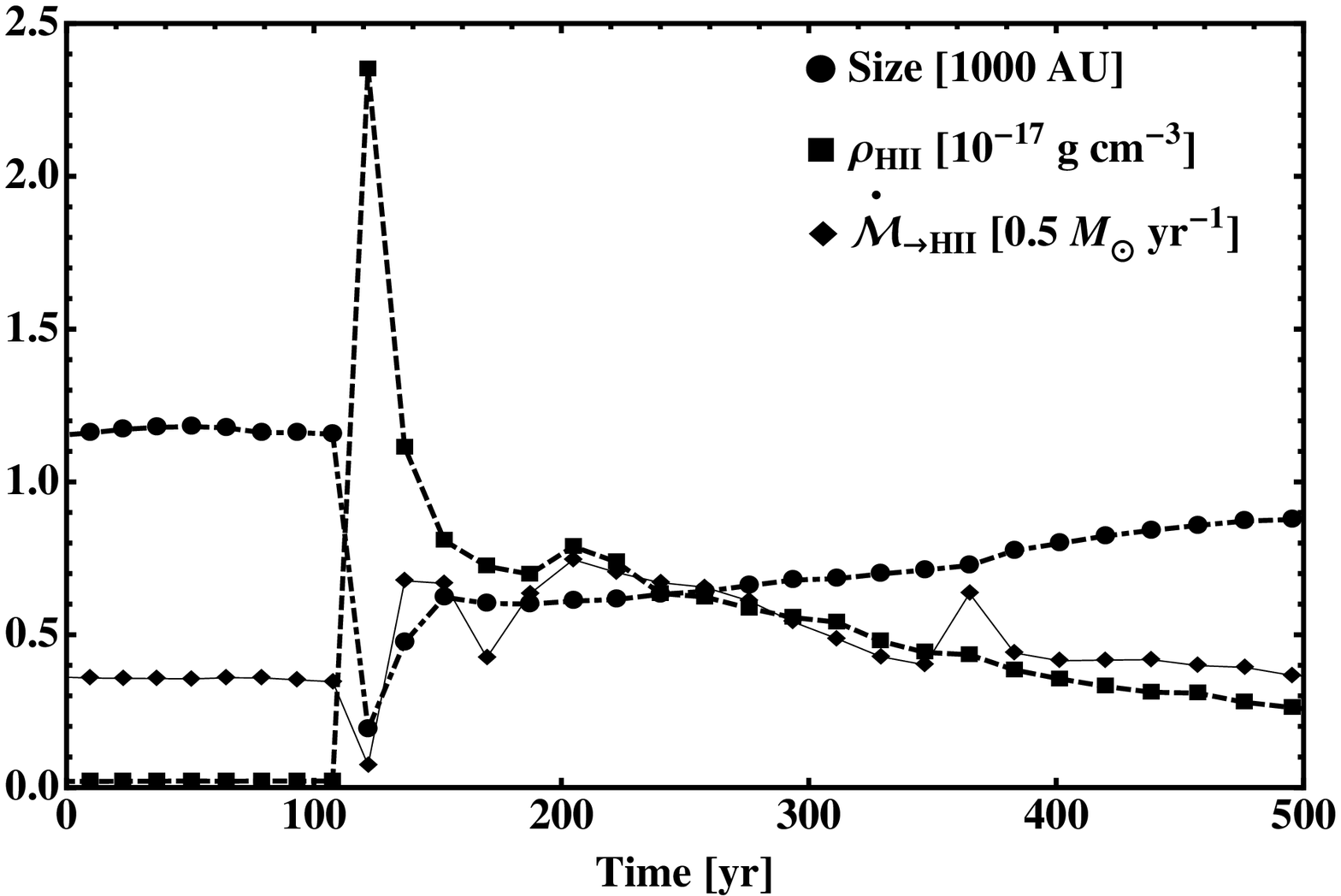}
\includegraphics[width=84mm]{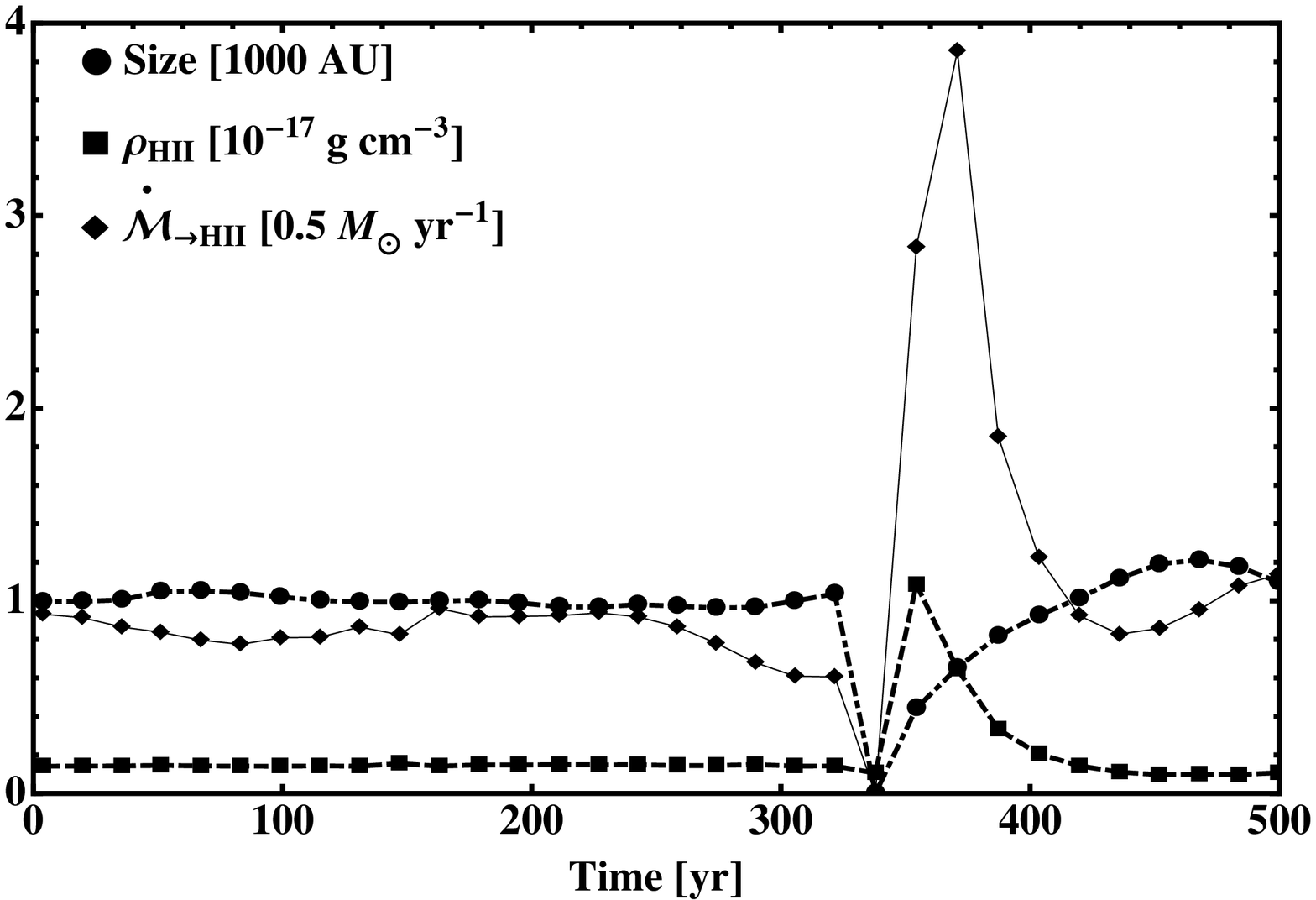}
\includegraphics[width=84mm]{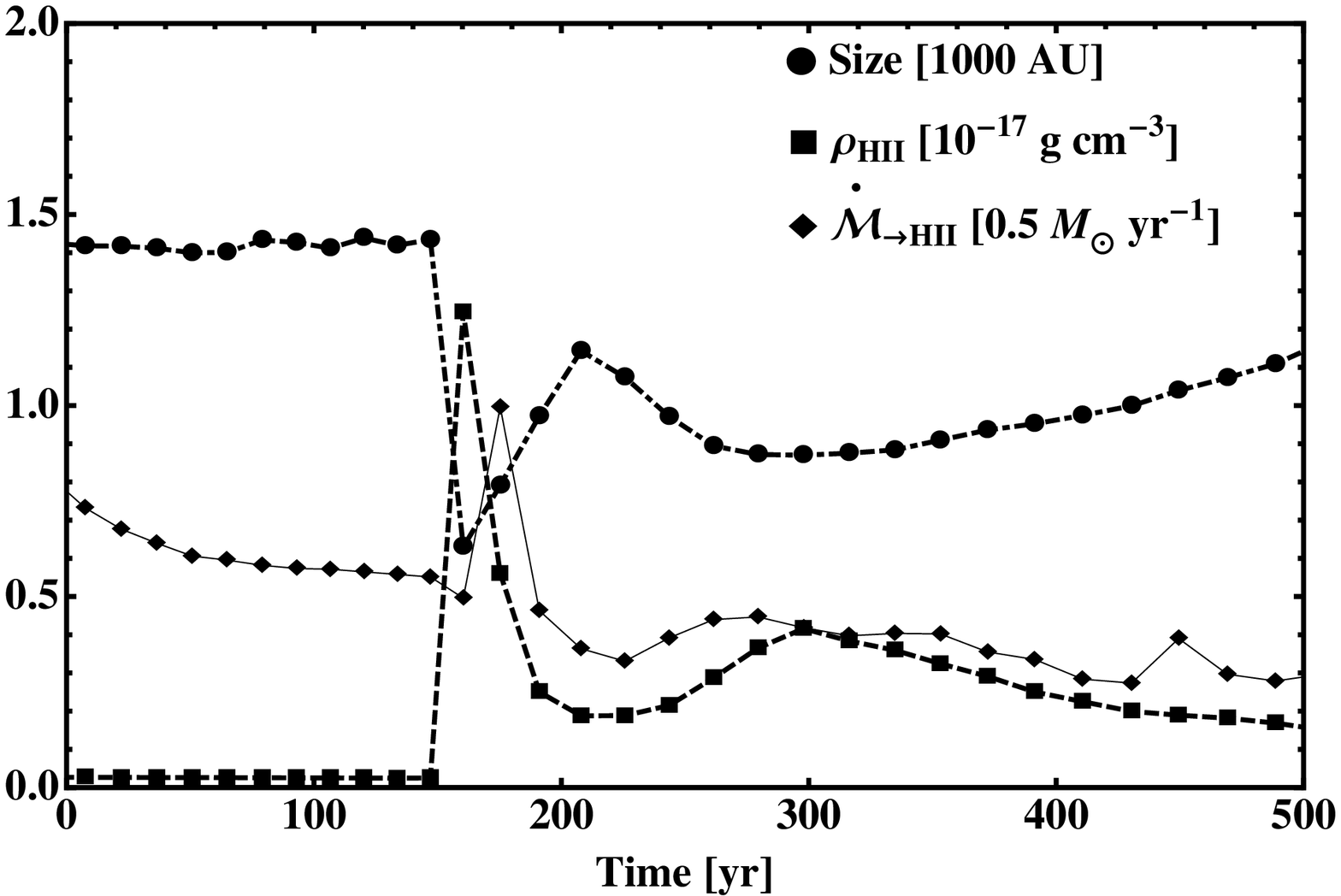}
\includegraphics[width=84mm]{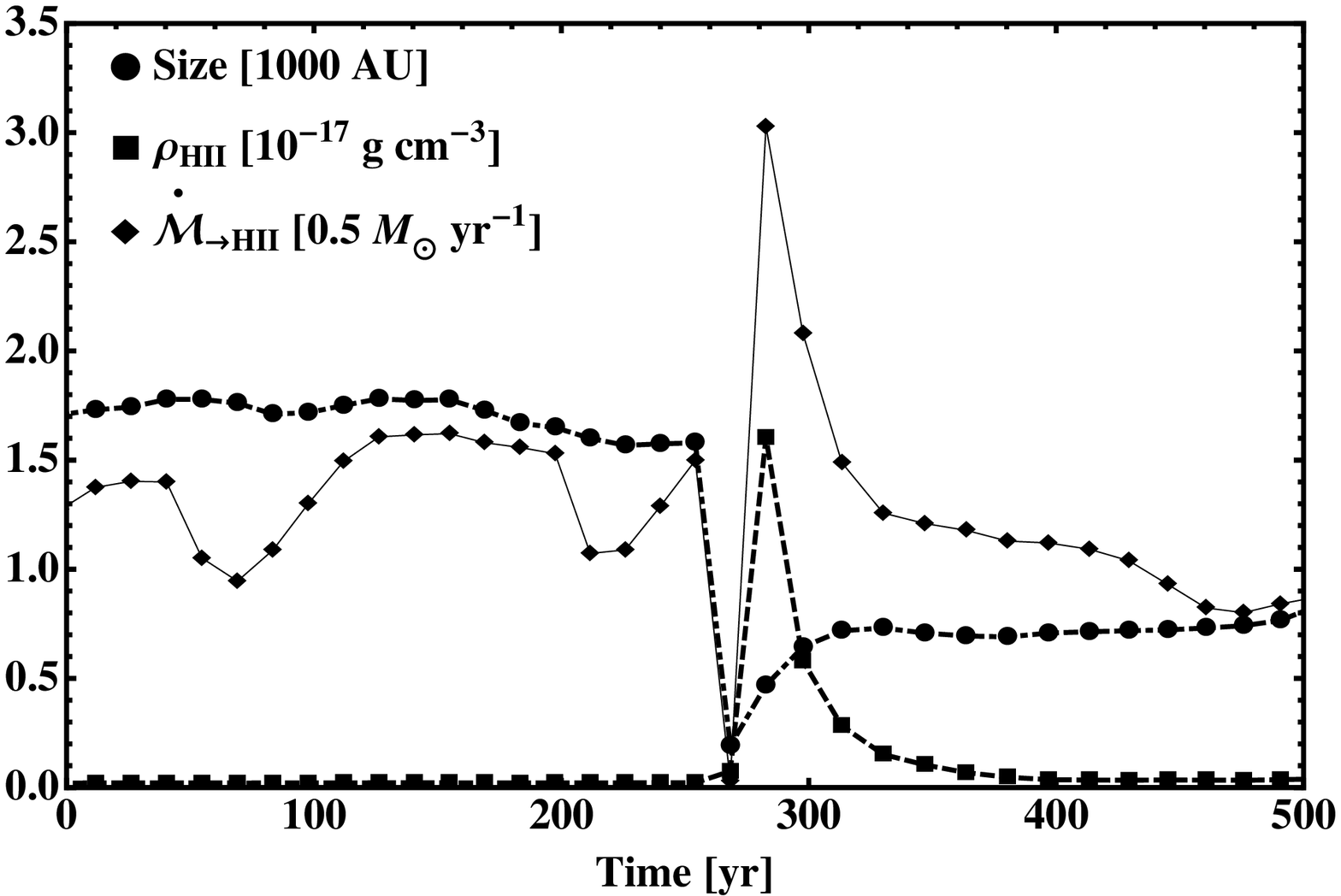}
\caption{Variation at high temporal resolution in Run B of the 
scale length $L_\mathrm{HII}$ (filled black circles) of the \HII region around the 
most massive sink particle, its average density $\rho_\mathrm{HII}$ (filled black squares), 
and its rate of ionization of neutral gas $\dot{M}_\mathrm{\rightarrow HII}$ 
(filled black diamonds).}
\label{fig14}
\end{figure}

\begin{figure}
\includegraphics[width=84mm]{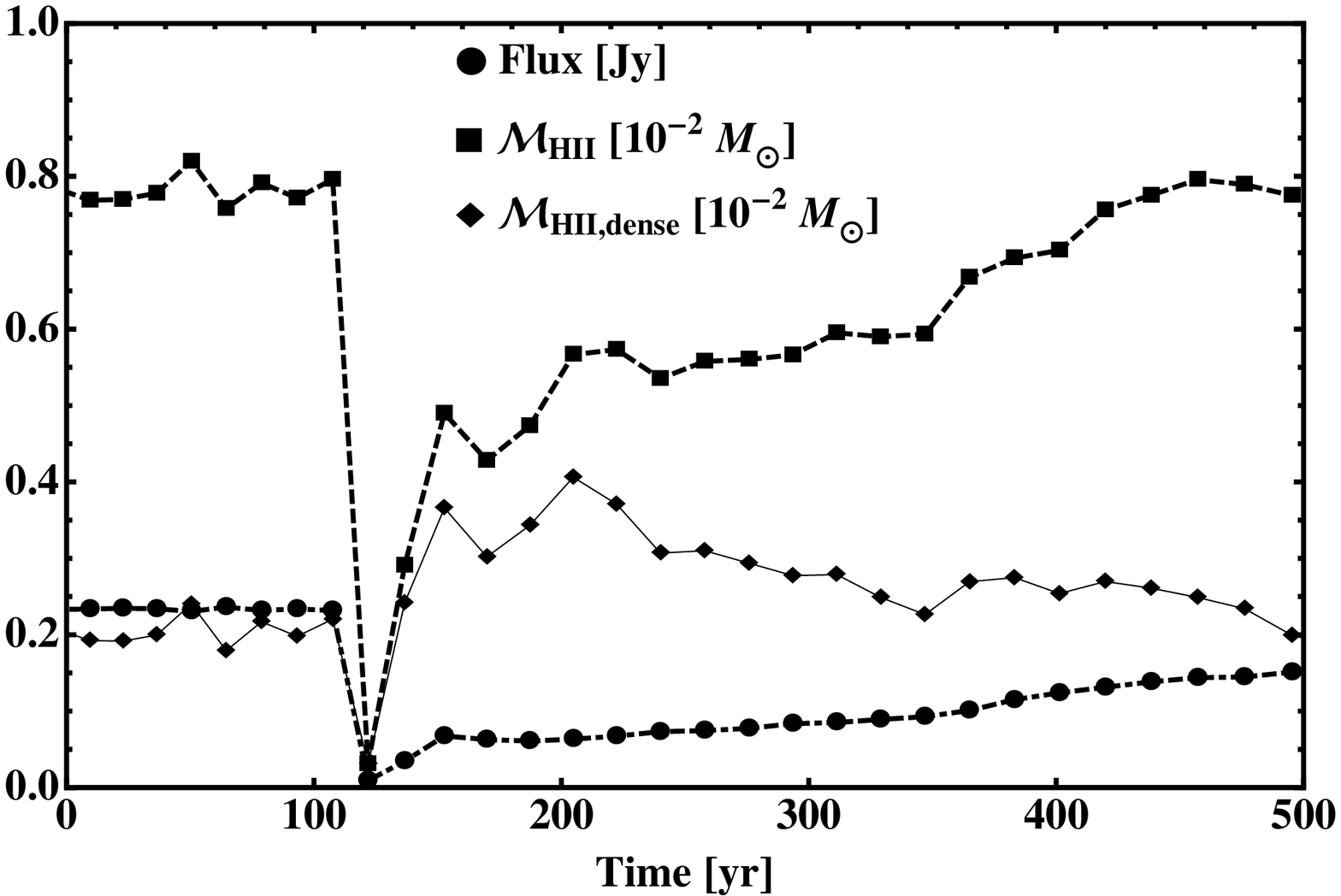}
\includegraphics[width=84mm]{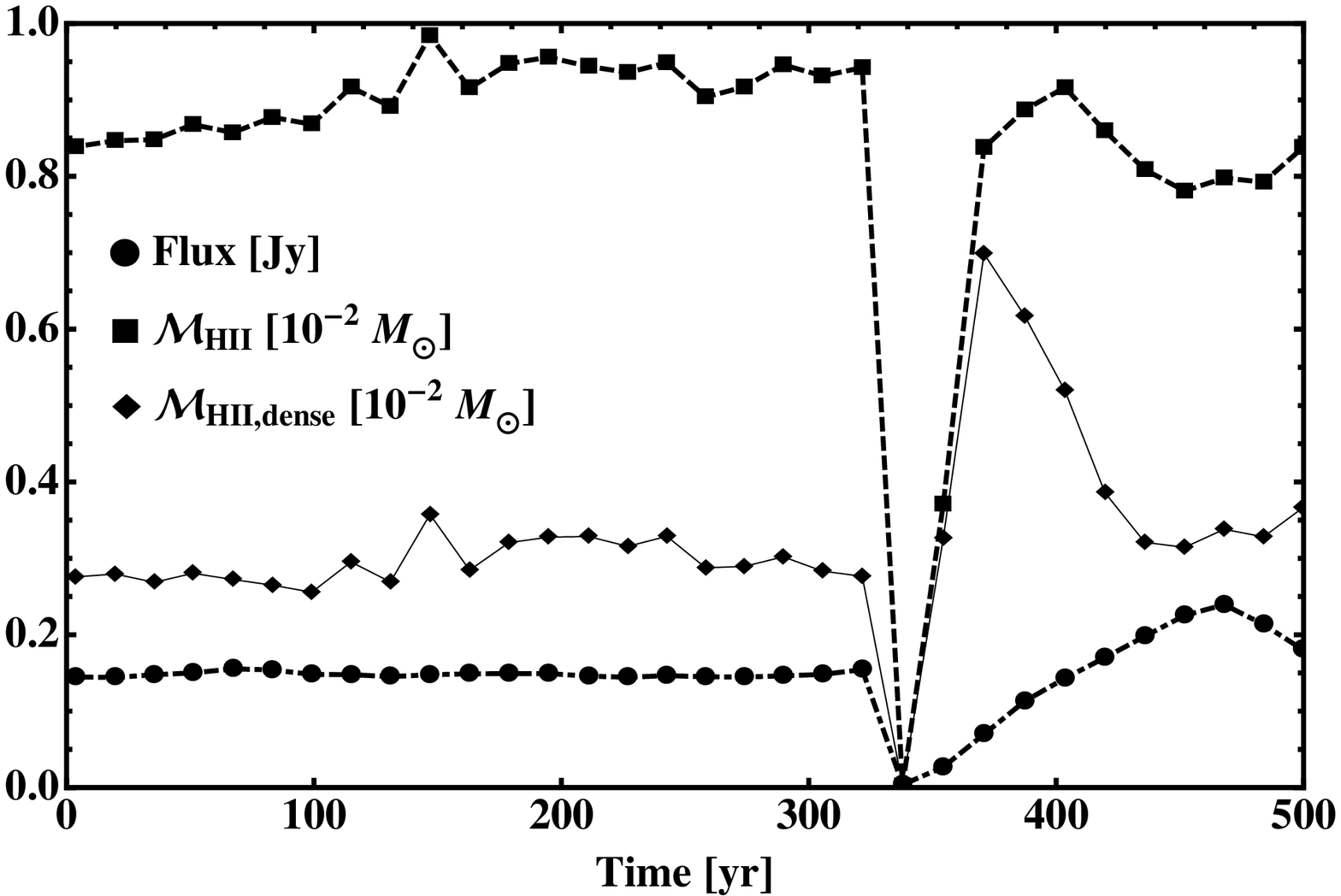}
\includegraphics[width=84mm]{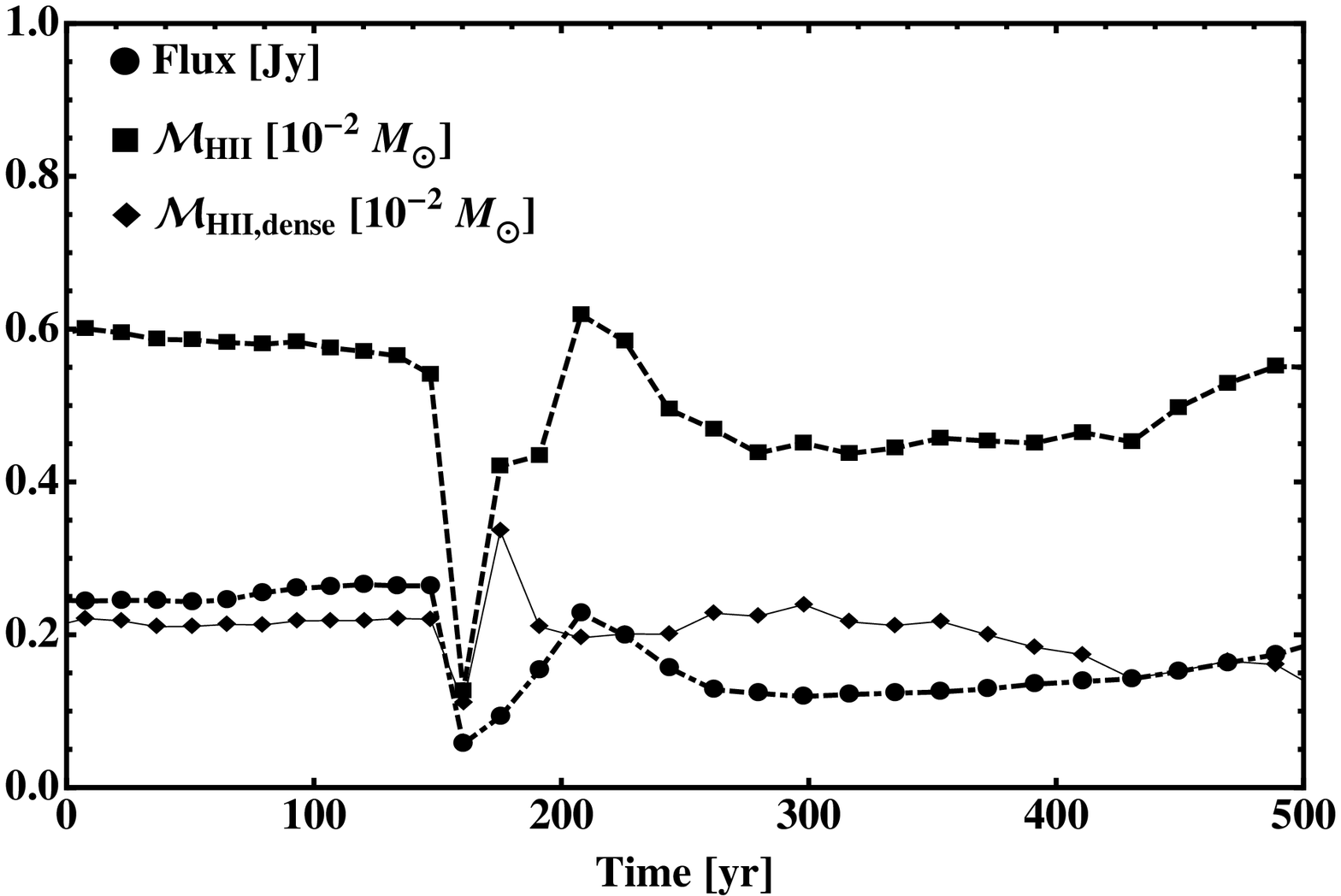}
\includegraphics[width=84mm]{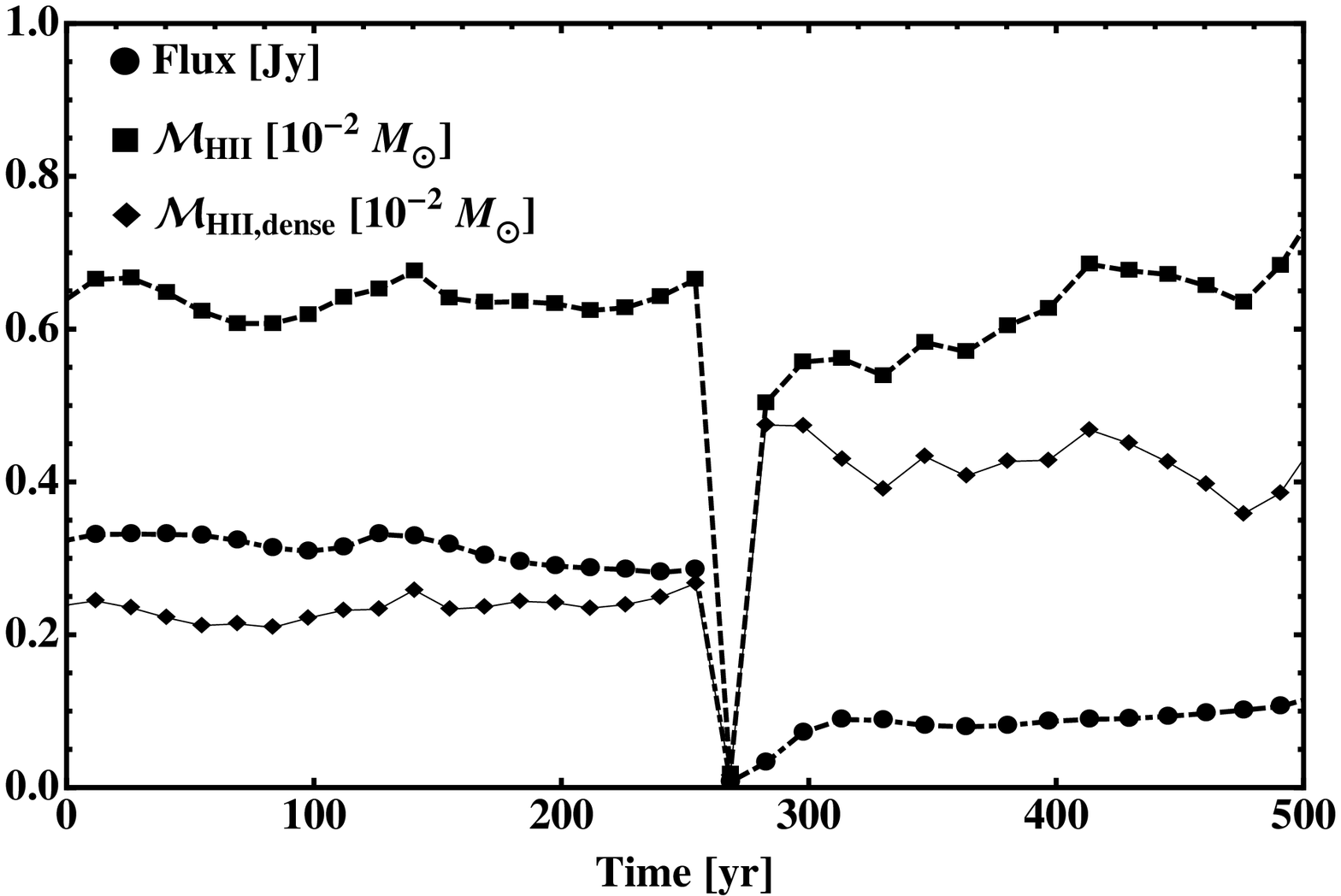}
\caption{Variation at high temporal resolution in Run B of the  
2-cm flux $S_\mathrm{2cm}$ (filled black circles) of the \HII region around the most massive 
sink particle, its ionized mass $M_\mathrm{HII}$ (filled black squares), and its 
dense-gas ionized mass $M_\mathrm{HII, dense}$ (filled black diamonds)}. 
\label{fig15}
\end{figure}

\subsection{Observational Signatures}

\indent

Time-variation effects 
are expected in the ionized gas but not in the molecular gas, since small clumps of molecular gas that 
become ionized and/or recombine have a much larger effect on the emission of the $< 1~M_\odot$ of 
ionized gas than on the tens to hundreds of $M_\odot$ of molecular gas in the accretion flow. 
One observational example is the HC \HII region G20.08N A, for which \cite{GM09} reports 
an ionized mass of $0.05~M_\odot$ and a mass of warm molecular gas in the inner 0.1 pc of 35 to 
95 $M_\odot$ \citep[another well studied example is G10.6--0.4, with more than $1000~M_\odot$ 
in the pc-scale molecular flow surrounding the \HII regions, e.g.,][and 
references therein]{Keto90,KW06,Liu11}. 
The \HII regions in the simulations also have masses of ionized gas in 
the range $10^{-3}$ to $10^{-1}~M_\odot$ while the stars are still accreting.

In sections 3.4 and 3.5 we have attempted to quantify the expected flux variations in the 
radio continuum of \HII regions for massive stars forming in isolation (Run A) and, more 
realistically, in clusters (Run B). The accretion flow in Run A is stronger than in Run B 
(actually the star in Run A never stops accreting, see Papers I and III), which leads 
to a brighter and more variable \HII region.  

For a given run, flux variation threshold, and 
time difference, positive variations are more likely to happen than negative ones, 
i.e., there is a constant struggle between the \HII region trying to expand and the 
surrounding neutral gas trying to confine it, with a statistical bias toward expansion. 

Monitoring \HII regions for thousands of years is not possible, but a few 
observations of rapid flux changes over timescales of $\sim10$ yr have been presented in the literature 
\citep{Acord98,FHR04, vdT05,GM08}. 
Comparing multi-epoch images made with radio-interferometers can be challenging: even if 
the absolute flux scale of the standard quasars is known to better than $2~\%$, slight 
differences in observational parameters between observations 
\citep[mainly the Fourier-space 
sampling, see][]{Perley89} make questionable any observed change smaller than $10~\%$. 
We have therefore measured 
the variation probabilities in the simulations at thresholds starting at $10~\%$. 
Considering this detection limit for the more realistic case of Run B, we have estimated that 
about $7~\%$ of UC and HC \HII regions observed at two epochs separated by about 10 years should have 
detectable flux increments, and that about $3~\%$ should have detectable 
decrements. In total, $\sim10~\%$ of \HII regions should have detectable flux variations 
in a period of 10 years. Dedicated observations of as many sources as possible are now needed 
to test this model. 

Our long timescale data can only be constrained by observational surveys, not by time monitoring. 
In Section 3.3 we have shown that the radio luminosities (i.e., distance-corrected flux) in 
the long-term evolution of the simulated \HII regions are consistent with major surveys, except 
for the most luminous \HII regions, in which likely accretion has stopped and which therefore do not 
correspond with our data. The hypothesis that a considerable fraction of observed UC and HC 
\HII regions may harbor stars that are still accreting material still needs more convincing 
evidence in addition to matching the model here presented. For most cases the dynamics of the surrounding 
molecular gas and of the ionized gas have not been studied at high angular resolution, and such 
studies in many sources are key to test this idea. From available observations,    
almost all of the massive star formation regions 
with signatures of active accretion and in which the mass of the protostar is  
estimated from dynamics to be $M_\star>20~\Msun$ have a relatively bright \HII region 
\citep[with at least $\sim 100$ mJy at short cm wavelenghts, e.g.,][]{Beltran07,GM09}. 
To our knowledge, the only clear exception is the recent report by \cite{Zap09} 
of an accreting protostar in W51 N with an estimated mass of $M_\star\sim 60 ~ \Msun$ 
and only 17 mJy at 7 mm. This object can be understood in the context of our simulations if 
it is in a quenched, faint state as currently observed.

\subsection{Caveats and limitations} 

\indent

As mentioned in Papers I, II, and III, the simulations here presented do not include the 
effects of stellar winds and magnetically-driven jets originating from within 100 AU, 
which would produce outflows that 
are more powerful than the purely pressure-driven outflows that appear in Runs A and B (Paper I). 

The inclusion of stellar winds and jets may affect the results 
presented in this paper only quantitatively. 
\cite{Peters11} have shown that magnetically 
driven outflows from radii beyond 100 AU do not stop 
accretion and even channel more material to the central most massive protostars. 
The simulations of \cite{Wang10} also indicate that collimated outflows may be 
an important regulator of  star formation by slowing the accretion rate but without 
impeding accretion. 
Observationally, molecular outflows tend to be less 
collimated for the more massive O-type protostars capable of producing \HII regions 
than for B-type protostars \citep{Arce07}. 
Regarding the radio-continuum, it is unknown if the free-free 
emission from the photoionized \HII regions produced by O-type protostars can coexist with   
the free-free emission from (partially) ionized, 
magnetically driven jets. Before the appearance of an \HII region, these 
jets are detected in protostars less massive than $\sim 15~\Msun$ 
\citep[e.g.,][]{Carrasco10}, and even though their radio emission also appears 
to be variable \citep[due to motions and interactions with the medium, e.g.,][]{Curiel06}, 
their typical centimeter flux is $\sim 1$ mJy, one to two orders of magnitude fainter 
than the typical flux of UC and HC \HII regions 
\citep[except maybe for the youngest gravitationally-trapped \HII regions, see][]{Keto03}. 
Therefore, the relative effect of any variation in a hypothetical radio jet 
should be small compared with the variations in the \HII region flux.  

A further limitation of this study is that accretion onto the protostars is not well resolved, 
since the maximum cell resolution (98 AU) corresponds to a scale of the order of the inner 
accretion disk (see also Paper I).

\section{Conclusions}

\indent

We performed an analysis of the radio-continuum variability in \HII 
regions that appear in the radiation-hydrodynamic simulations of massive-star formation 
presented in Paper I. The ultimate fate of ultracompact and hypercompact \HII regions is 
to expand, but during their evolution they flicker due to the complex interplay of the inner 
ionized gas and the outer 
neutral gas. The radio-luminosities of the \HII regions formed by the accreting protostars in 
our simulations are in agreement with those of observational radio surveys, except for the 
most luminous of the observed \HII regions. We show that \HII regions are highly variable in 
all timescales from 10 to $10^4$ yr, and estimate that at least $10 ~\%$ of observed ultracompact 
and hypercompact \HII regions should exhibit flux variations larger than $10 ~\%$ for time intervals 
longer than about 10 yr.

\section*{Acknowledgments}

\indent

The authors acknowledge the referee for a report that helped to clarify the main aspects 
of this paper. R.G.M. thanks Luis F. Rodr\'iguez for comments on a draft of the paper. 
R.G.M. acknowledges support from the SAO and ASIAA  through an SMA predoctoral fellowship.  
T.P. is a Fellow of the Baden-W\"{u}rttemberg Stiftung funded by their program International
Collaboration II (grant P-LS-SPII/18). T.P. also acknowledges support from an Annette Kade Fellowship 
for his visit to the AMNH and a Visiting Scientist Award of the SAO.
R.S.K.\ acknowledges financial support from the Baden-W\"{u}rttemberg Stiftung
via their program International Collaboration II (grant P-LS-SPII/18) and from the German
Bundesministerium f\"{u}r Bildung und Forschung via the ASTRONET project STAR FORMAT (grant 05A09VHA).
R.S.K. furthermore gives
thanks for subsidies from the Deutsche Forschungsgemeinschaft (DFG) under
grants no.\ KL 1358/1, KL 1358/4, KL 1359/5, KL 1358/10, and KL 1358/11, as well as from a Frontier
grant of Heidelberg University sponsored by the German Excellence Initiative.
M.-M.M.L. was partly supported by NSF grant AST 08-35734.
R.B. is funded by the DFG via the Emmy-Noether grant BA 3706/1-1.
We acknowledge computing time at the Leibniz-Rechenzentrum in Garching (Germany), the NSF-supported
Texas Advanced Computing Center (USA), and at J\"ulich Supercomputing Centre (Germany). The FLASH code
was in part developed by the DOE-supported Alliances Center for Astrophysical Thermonuclear
Flashes (ASCI) at the University of Chicago.

\end{document}